\documentclass[prd,eqsecnum,twocolumn,nofootinbib]{revtex4-1}
\usepackage{amsfonts}
\usepackage{amsmath}
\usepackage{amssymb}
\usepackage{amsthm}
\usepackage{bm}
\usepackage{dcolumn}
\usepackage{float}
\usepackage{epsfig}
\usepackage{graphicx}
\usepackage{graphics}
\usepackage[latin1]{inputenc}
\usepackage{latexsym}
\usepackage{rotating}
\begin{document}

\title{Exciting black hole modes via misaligned coalescences: I. Inspiral, transition, and plunge trajectories using a generalized Ori-Thorne procedure}

\author{Anuj Apte} \affiliation{Department of Physics and MIT Kavli Institute, Massachusetts Institute of Technology, Cambridge, MA 02139}

\author{Scott A.\ Hughes} \affiliation{Department of Physics and MIT Kavli Institute, Massachusetts Institute of Technology, Cambridge, MA 02139}

\begin{abstract}
The last gravitational waves emitted in the coalescence of two black holes are quasi-normal ringing modes of the merged remnant.  In general relativity, the mass and the spin of the remnant black hole uniquely determine the frequency and damping time of each radiated mode.  The amplitudes of these modes are determined by the mass ratio of the system and the geometry of the coalescence.  This paper is part I of an analysis that aims to compute the ``excitation factors'' associated with misaligned binary black hole coalescence.  To simplify the analysis, we consider a large mass ratio system consisting of a non-spinning body of mass $\mu$ that inspirals on a quasi-circular trajectory into a Kerr black hole of mass $M$ and spin parameter $a$, with $\mu/M \ll 1$.  Our goal is to understand how different modes are excited as a function of the black hole spin $a$ and an angle $I$ which characterizes the misalignment of the orbit with the black hole's spin axis.  Though the large mass ratio limit does not describe the binaries that are being observed by gravitational-wave detectors today, this limit makes it possible to quickly and easily explore the binary parameter space, and to develop insight into how the system's late ringing waves depend on the binary's geometry.  In this first analysis, we develop the worldline which the small body follows as it inspirals and then plunges into the large black hole.  Our analysis generalizes earlier work by Ori and Thorne to describe how a non-equatorial circular inspiral transitions into a plunging trajectory that falls into the black hole.  The worldlines which we develop here are used in part II as input to a time-domain black hole perturbation solver.  This solver computes the gravitational waves generated by such inspirals and plunges, making it possible to characterize the modes which the coalescence excites.
\end{abstract}

\maketitle

\section{Introduction}
\label{sec:intro}

\subsection{Background and motivation}
\label{sec:motivate}

Since 14 September 2015 until this text was written, the LIGO-Virgo Collaboration (LVC) has measured gravitational waves (GWs) from over a dozen binary black hole (BBH) coalescences (Refs.\ {\cite{gw150914, gw151226, gw170104, gw170814, gw170608, gwtc1}} present events that have been studied in depth; several recent detections have been announced via the Open Alerts\footnote{https://gcn.gsfc.nasa.gov/lvc\_events.html} that accompany the 3rd observing run of the LVC).  BBH coalescences have proven so far to be the most frequently measured sources for ground-based GW detectors.

A surprise associated with these events is that many of these black holes are substantially more massive than had been expected based on measurements of stellar-mass black hole prior to the first LVC discovery.  In the recently published catalog of events from the first two observing runs {\cite{gwtc1}}, the majority of events analyzed to date involve black holes with $M \gtrsim 20\,M_\odot$.  Because the frequency spectrum of a GW source scales inversely with the source's mass, this means that the peak sensitivity of detectors like LIGO and Virgo for such sources corresponds to the late ``merger" and ``ringdown'' waves, emitted when the two black holes merge into a single body and settle down to the stationary Kerr solution.  The early ``inspiral'' waves for these sources, which encode detailed information about the masses and spins of the binary's constituents, are at low frequencies for which detectors are less sensitive.  For high mass systems, we will not get the benefit of the many inspiral wave cycles that encode the widely separated binary's properties.

Although the inspiral waves are less informative for high mass systems, such binaries generate perhaps an ideal spectrum of merger and ringdown waves for ground-based GW detectors.  Especially as detector upgrades improve our ability to measure these waves {\cite{et1, et2, ce1, ce2, voyager, LIGOLF}}, BBH merger and ringdown waves will be measured with high fidelity, and will comprise an important component of the measured GW catalog in coming years.  They are also certain to be important components of the catalog that will be measured by the space-based detector LISA {\cite{LISA}} expected to launch in the early 2030s, which will measure the low-frequency GWs from processes involving black holes with $M \sim \mbox{several}\times 10^4\,M_\odot$ -- $\mbox{several}\times 10^7\,M_\odot$.

What information is encoded in those waves?  This is simplest to answer for the ringdown waves, for which we have good analytic understanding.  Each ringing mode has a waveform which is simply a damped sinusoid.  From the measured frequencies and damping times of the modes, we will be able to determine the masses and spins of the remnant black holes {\cite{qnr1}}.  If multiple modes can be measured, consistency of these modes will make it possible to check the validity of the Kerr metric and to make other interesting tests of strong-field gravity {\cite{qnr2}}.

What can we learn from measuring the modes' amplitudes?  A given coalescence will excite multiple modes of the remnant black hole.  The relative amplitude of the modes will depend on the geometry of the final plunge and coalescence of the binary's members.  For example, the $\ell = 2$, $m = 2$ mode is likely to be the most strongly excited mode for an equatorial coalescence, but a highly inclined coalescence will excite the different $\ell = 2$ modes in a more ecumenical manner.  Might we be able to learn some of the properties of the progenitor binary by measuring multiple ringdown modes?

To answer this, we must compute how different modes are excited as a function of binary properties.  These ``excitation factors'' will certainly depend on the black hole's spin $a$ as well as an angle $I$ describing the orientation of the black hole's spin axis and the smaller body's orbital plane.  The modes may also depend on ``accidental'' phases: orbits which share the same orbital plane but end their plunges at different values of the polar angle $\theta$ may excite different mode mixtures.  A detailed analysis is needed to assess whether the late merger and ringdown modes can be used to learn about the progenitor binary.

\subsection{Binary coalescence in the large mass-ratio limit}
\label{sec:binarycoal}

Our goal is to begin this detailed analysis.  We do so in the large mass-ratio limit, taking the binary to be the exact Kerr solution of general relativity plus a smaller body which perturbs its spacetime.  We strongly emphasize at this point that the goal of our analysis is {\it not} to produce templates which could be used for present gravitational-wave data analysis, but rather to explore the physics of mode excitation.  The large mass-ratio limit is an excellent tool for such exploration, since it lets us easily vary binary parameters and rapidly compute the waveforms corresponding to each parameter choice.

The only astrophysical source to which this limit applies are the ``extreme mass-ratio inspirals," or ``EMRIs" {\cite{emris}}, important sources for space-based detectors like LISA.  However, insights from the large mass-ratio regime have proven useful much more generally, even in the analysis of nearly equal mass binaries.  For example, many quantities can be computed to very high precision in the large mass-ratio limit.  The behavior of GW fluxes and the self force in this regime has played an important role in refining the effective one-body approach to binary dynamics {\cite{eob_sf, tbhk13}}, essentially by providing a precisely computed asymptotic regime to which all other quantities must limit.  Nonetheless, calculations in the comparable mass regime are needed in order to understand what large mass-ratio insights carry over.  Our hope is that the promise of the results we report will motivate further investigation in this vein.

Our goal is to use the large mass-ratio limit to set up an easily parameterized binary that will allow us to explore the dynamics of black hole mode excitation.  We will begin by considering a small body that is moving on an initially circular orbit of a Kerr black hole.  GW emission drives the small body to spiral through a sequence of circular orbits of ever smaller radius.  Eventually, the small body reaches the vicinity of the ``innermost stable circular orbit" or ``ISCO," beyond which circular orbits are no longer stable against small disturbances.  The body then follows a plunging trajectory which crosses the event horizon in finite proper time.

Following this prescription of slow inspiral followed by a transition and plunge, it is straightforward to construct the worldline that a small body follows as it moves through spacetime into the larger black hole.  We then use this worldline as input to a time-domain black hole perturbation theory (BHPT) code {\cite{td1,td2,td3,td4}}.  This code computes the GWs generated by the small body as it follows this trajectory.  The behavior of the BHPT source term as the small body approaches the horizon guarantees that the final waveform cycles the code computes are the quasi-normal modes generated by the coalescence.

In summary, the goal of our analysis is to generate worldlines corresponding to a small body that inspirals and plunges into a Kerr black hole, to compute the GWs generated by that inspiral and plunge, and then to characterize the different quasi-normal modes that are thereby excited.  By considering a range of worldlines corresponding to different parameter choices (varying, for example, the spin of the large black hole, or the misalignment between the orbit and the black hole's spin axis), we aim to understand how the spectrum of late ringing modes varies as a function of the binary's properties at merger.

\subsection{The transition between inspiral and plunge}
\label{sec:intro_trans}

We split this task into two parts.  In this paper, we focus on how to compute the worldline that the small body follows as it spirals into and then plunges into the black hole.  In a companion paper {\cite{lkah}}, we describe how we use that worldline to compute the corresponding GWs (much of this already has appeared in the literature; see Refs.\ {\cite{td1,td2}}), and how to characterize the ringdown modes which the plunge and merger excites.

A major focus of this paper is to describe how the small body transitions from a slowly evolving, nearly circular geodesic to a plunging geodesic which carries the small body into the black hole.  This epoch of the small body's motion must be treated with some care.  For much of the binary's history, the small body can be regarded as being on a circular, geodesic orbit.  Being on such an orbit means that the small body sits at the extremum of a potential-like function which characterizes black hole orbits.  Due to GW emission, this potential's properties adiabatically change, moving the extremum to smaller radius.  As long as the orbit is far from the ISCO, the rate at which the extremum moves inward is slow enough that the smaller body can ``keep up": the curvature of the potential provides a restoring force which keeps the small body on a circular geodesic orbit.  Far from the ISCO, inspiral is thus accurately described as the small body moving through a sequence of geodesics, with GW emission determining the sequence that is followed.

As the small body approaches the ISCO, the situation changes.  The potential flattens, losing its extremum altogether at the ISCO.  The restoring force which kept the small body on a circular orbit becomes weaker as the potential becomes flatter.  Eventually, the small body can no longer keep up with the evolving potential, and the motion ceases to be geodesic.  It falls ever more rapidly toward the larger black hole, asymptotically approaching an infalling geodesic which plunges into the black hole's event horizon.

For equatorial orbits, work over the past nearly two decades using the ``effective one-body" (EOB) framework has developed a very mature set of tools for modeling this final transition and plunge {\cite{bd00,ndt07,dn07,bbhkop12,tbkh14}}.  EOB waveforms calibrated to the output of numerical relativity simulations build in this behavior {\cite{taracchinietal14}}, and have played an important role in the discovery and interpretation of GWs.  Ideally, we would use EOB equations of motion in the large mass-ratio limit to develop inspiral and plunge worldlines, much as was done in Refs.\ {\cite{ndt07,dn07,bbhkop12,tbkh14}}.  An EOB formalism appropriate for binaries with one or more spins misaligned to the orbit has long existed {\cite{bcd05}}, and forms the foundation for ongoing work to model gravitational waveforms for such systems {\cite{panetal14,btb17}}.  This formalism is adapted to the post-Newtonian formulation of the motion of spinning bodies; some work must be done to match it to the Kerr geodesic description that is appropriate in the large mass-ratio limit (carefully translating, for example, the notion of angular momentum used in Refs.\ {\cite{bcd05,panetal14,btb17}} to the axial angular momentum $L_z$ and Carter constant $Q$ used to describe Kerr orbits).

In lieu of developing a match of the ``precessing plunge" EOB model to strong-field Kerr geodesic orbits, we will instead use a framework that is similar to the equations of motion that were developed by Ori and Thorne {\cite{ot00}} (hereafter OT00) for equatorial orbits.  OT00 developed equations of motion for the non-geodesic epoch following inspiral by examining how the geodesic equation's character changes as the potential governing radial motion flattens out, taking into account the leading impact of radiative backreaction.  The key equation derived by OT00 governing the motion from the end of inspiral to plunge turns out to be identical (modulo minor differences in some definitions) to an analogous equation found in the EOB framework [compare Eq.\ (3.22) of OT00 with Eq.\ (4.38) of Ref.\ {\cite{bd00}}], at least in the large mass-ratio limit.  OT00 and the EOB framework thus produce the same motion, at least over a certain regime.  However, as we'll discuss in Sec.\ {\ref{sec:plunge}}, the OT00 procedure has some unsatisfactory arbitrariness which is not well modeled.  This motivates work to enable the match of the precessing EOB model to strong-field Kerr geodesics.

Setting aside for now these concerns, we thus seek to generalize OT00 from equatorial circular orbits to circular orbits of arbitrary spin-orbit alignment.  Such an analysis was done by Sundararajan {\cite{pranesh}}.  As we will describe in Sec.\ {\ref{sec:GOT}}, we have found that Sundararajan's analysis is unfortunately flawed.  By using {\it Mino time} {\cite{mino03}}, a parameterization of strong-field Kerr orbits which separates the radial and polar motions of Kerr black hole orbits, we show that a very simple generalization of OT00 can be used to describe the transition for circular orbits of arbitrary orientation.  Constructing a worldline for arbitrary spin-orbit configurations is then essentially no more difficult for inclined inspiral and plunge than it is for the equatorial case.

\subsection{Organization of this paper}
\label{sec:org}

The remainder of this paper is organized as follows.  We begin by reviewing Kerr black hole orbits in Sec.\ {\ref{sec:kerr}}.  Since the geodesics of the Kerr spacetime play an important foundational role in our analysis, it is useful to have their properties and certain critical results at our fingertips.  In Sec.\ {\ref{sec:transitionEOM}}, we next develop the equations which govern the motion of a small body through the transition.  Section {\ref{sec:OT}} presents a synopsis of the OT00 calculation, showing how to model the transition from inspiral to plunge for equatorial orbit configurations.  We then generalize the OT00 analysis in Sec.\ {\ref{sec:GOT}} to model the transition for a binary of any orbit inclination.  Section {\ref{sec:plunge}} concludes this section by describing how we model the small body's motion once it is best described as a free-fall plunge into the larger black hole.

In Sec.\ {\ref{sec:worldline}}, we show how to stitch the sequence of circular geodesics describing adiabatic inspiral to the transition and plunge, thereby constructing the worldline which the small body follows as it falls into the larger black hole.  A critical step in this process is to choose when we end the adiabatic inspiral and begin the transition, and when we end transition and begin the plunge.  We argue that internal self consistency demands that transition begin within a certain fairly narrowly defined interval.  The interval we find is consistent with a more heuristic picture of the transition that we develop in Appendix {\ref{app:heuristic}}.  We likewise argue that internal self consistency demands a fairly narrow window in which to end the transition.  In this section, we also highlight the need to modify a particular step from OT00 slightly in order to guarantee that certain quantities evolve smoothly as inspiral ends and the transition begins.

Examples of our results are shown in Sec.\ {\ref{sec:results}}.  An important feature of our worldlines (highlighted in earlier work, notably Refs.\ {\cite{td3,ndt07,dn07}}) is that the small body's motion freezes to the generators of the event horizon at late times as seen by distant observers.  When coupled to a time-domain BHPT code, this has the important result that the wave equation's source term redshifts to zero at late times.  The late-time radiation from the system thus consists of quasi-normal modes of the larger black hole, produced in a way that, by construction, is phase coherent with the earlier wave train from the coalescence.

In the presentation of our results, we examine how robust the worldlines we develop are to the {\it ad hoc} choices that we must make regarding the start and end of the transition domain.  Our results depend very little on when we choose the transition to begin, barely changing as we vary our start parameter across the acceptable range of transition start times.  However, we find that the worldlines do vary quite a bit as a function of our transition end time.  Fortunately, a companion analysis {\cite{lkah}} shows that this dependence does not have a detrimental impact on our ability to study the question driving this research.  Although the worldlines depend on when we change from transition to plunge, we find that this choice has unimportant consequences for understanding the excitation of the black hole's ringdown modes.

In Sec.\ {\ref{sec:conclude}}, we summarize our conclusions and briefly describe the work which is presented in our companion analysis {\cite{lkah}}.  Throughout this paper, we use units in which $G = 1 = c$.  We typically work in Boyer-Lindquist coordinates describing a black hole of mass $M$ with spin parameter $a \equiv S/M$, where $S$ is the hole's spin angular momentum.  The orbiting body has mass $\mu$; we define $\eta \equiv \mu/M$.

\section{Important properties of Kerr orbits}
\label{sec:kerr}

\subsection{Generalities}
\label{sec:kerrgen}

As background to our discussion, we first summarize key equations describing geodesic orbits of Kerr black holes.  In first order form, the geodesic motion of a body in the Kerr spacetime in the Boyer-Lindquist coordinates $(r,\theta,\phi,t)$ is governed by the equations\begin{eqnarray}
\Sigma^2\left(\frac{dr}{d\tau}\right)^2 &=& \left[E(r^2+a^2) - a L_z\right]^2
\nonumber\\
& & - \Delta\left[r^2 + (L_z - a E)^2 + Q\right]
\nonumber\\
&\equiv& R(r)\;,\label{eq:rdot}\\
\Sigma^2\left(\frac{d\theta}{d\tau}\right)^2 &=& Q - L_z^2 \cot^2\theta - a^2\cos^2\theta[1 - E^2]
\nonumber\\
&\equiv&\Theta(\theta)\;,\label{eq:thetadot}\\
\Sigma\left(\frac{d\phi}{d\tau}\right) &=& \csc^2\theta L_z + \frac{2 M r a E}{\Delta} - \frac{a^2L_z}{\Delta}
\nonumber\\
&\equiv&\Phi(r,\theta)\;,\label{eq:phidot}\\
\Sigma\left(\frac{dt}{d\tau}\right) &=& E\left[\frac{(r^2+a^2)^2}{\Delta} - a^2\sin^2\theta\right] - \frac{2 M r a L_z}{\Delta}
\nonumber\\
&\equiv& T(r,\theta)\label{eq:tdot}\;,
\end{eqnarray}
where
\begin{equation}
\Delta \equiv r^2 - 2Mr + a^2\;,\quad
\Sigma \equiv r^2 + a^2\cos^2\theta
\end{equation}
[see Ref.\ {\cite{mtw}}, Eqs.\ (33.32a)--(33.32d)].  The quantity $E$ is the orbit's energy (per unit $\mu$, the mass of the orbiting body), $L_z$ is the axial angular momentum (per unit $\mu$), and $Q$ is the orbit's Carter constant (per unit $\mu^2$).  These quantities are constants along a particular geodesic orbit.  Up to initial conditions, an orbit is specified by choosing values for $E$, $L_z$, and $Q$.

Equations (\ref{eq:rdot})--(\ref{eq:tdot}) use proper time $\tau$ as the independent parameter along the geodesic.  Another time parameter which is very useful for studying strong-field Kerr black hole orbits is $\lambda$, defined by $d\lambda = d\tau/\Sigma$.  The geodesic equations parameterized in this way are
\begin{eqnarray}
\left(\frac{dr}{d\lambda}\right)^2 = R(r)\;,
&\qquad&
\left(\frac{d\theta}{d\lambda}\right)^2 = \Theta(\theta)\;,
\nonumber\\
\frac{d\phi}{d\lambda} = \Phi(r,\theta)\;,
&\qquad&
\frac{dt}{d\lambda} = T(r,\theta)\;.
\label{eq:geodesiceqnsMino}
\end{eqnarray}
By using $\lambda$ as our time parameter, the $r$ and $\theta$ coordinate motions separate.  The parameter $\lambda$ is often called ``Mino time,'' following Mino's use of it to untangle these coordinate motions {\cite{mino03}}.

\subsection{Circular orbits}
\label{sec:kerrcirc}

In our analysis, we focus on ``circular'' orbits: orbits which have constant Boyer-Lindquist radius $r$.  Such orbits are defined by enforcing the conditions $R = 0$, $R' = 0$ (where $' \equiv \partial/\partial r$).  Orbits satisfying these circularity conditions have $dr/d\tau = 0$ (or $dr/d\lambda = 0$) at all times.  Enforcing circularity yields a two-parameter orbit family.  We will take the parameters to be the orbital radius\footnote{We write a subscript ``o'' on the orbital radius to contrast the orbit parameter with the general radial coordinate $r$.} $r_{\rm o}$ and an inclination angle $I$.  Orbits with $I = 0^\circ$ are confined to the equatorial plane (polar angle $\theta = \pi/2$ for all time), and have $L_z > 0$ (i.e., are prograde); those with $I = 180^\circ$ are also equatorial, but have $L_z < 0$.  For general $I$, $\theta$ oscillates between
\begin{eqnarray}
\theta_{\rm min} &=& {\rm sgn}(L_z)\times\left(\frac{\pi}{2} -
I\right)\;,\qquad{\rm and}
\nonumber\\
\theta_{\rm max} &=& \pi - \theta_{\rm min}\;.
\label{eq:thetaminmaxdefs}
\end{eqnarray}
Once $r_{\rm o}$ and $I$ are selected, $E$, $L_z$, and $Q$ are found by solving $R(r_{\rm o}) = 0$, $R'(r_{\rm o}) = 0$, $\Theta(\theta_{\rm min}) = 0$.

Begin by considering equatorial orbits, for which $\theta = \pi/2$ at all times.  Using Eq.\ (\ref{eq:thetadot}), we see this requires $Q^{\rm eq} = 0$.  Solving $R = 0$, $R' = 0$ with $Q = 0$ yields {\cite{bpt72}}
\begin{eqnarray}
E^{\rm eq} &=& \frac{1 - 2v^2 \pm qv^3}{\sqrt{1 - 3v^2 \pm 2qv^3}}\;,
\label{eq:E_eq}\\
L^{\rm eq}_z &=& \pm r_{\rm o} v\frac{1 \mp 2qv^3 + q^2v^4}{\sqrt{1 -
    3v^2 \pm 2qv^3}}\;.
\label{eq:Lz_eq}
\end{eqnarray}
In these expressions and those which follow, the upper sign refers to prograde orbits ($I = 0^\circ$), and the lower to retrograde ($I = 180^\circ$).  We have introduced $v \equiv \sqrt{M/r_{\rm o}}$ and $q \equiv a/M$.

Generalizing to non-equatorial orbits is straightforward; we refer the reader to Appendix B of Schmidt {\cite{schmidt}} for detailed discussion and formulas.  Software implementing these formulas can be found at the Black Hole Perturbation Toolkit {\cite{bhpt}}.  The results are simple in the Schwarzschild limit, $a = 0$:
\begin{eqnarray}
E^{\rm Schw} &=& \frac{1 - 2v^2}{\sqrt{1 - 3v^2}}\;,
\nonumber\\
L^{\rm Schw}_z &=& \cos I \frac{r_{\rm o} v}{\sqrt{1 - 3v^2}}\;,
\nonumber\\
Q^{\rm Schw} &=& \sin^2I \frac{(r_{\rm o} v)^2}{1 - 3v^2}\;.
\end{eqnarray}
For general $a$, the analogous formulas are complicated, so we do not show them explicitly here.  All of these quantities vary monotically as functions of $I$ at fixed $r_{\rm o}$ and $a$.  Once all of these quantities are known, it is very useful to reparameterize the $\theta$ motion:
\begin{equation}
\cos\theta = \sin I\cos\left(\chi + \chi_0\right)\;.
\label{eq:chidef}
\end{equation}
The parameter $\chi_0$ is essentially a starting phase for the polar angle; as we'll describe in later sections, we use it to control the value of $\theta$ at which members of a family of worldlines with the same $I$ enter and complete the plunge.  Using Eqs.\ (\ref{eq:thetadot}) and (\ref{eq:chidef}), it is simple to show that $\chi$ is governed by the equation
\begin{equation}
\frac{d\chi}{d\lambda} = \sqrt{\beta(z_+ - \cos^2\theta)}\;,
\label{eq:dchidlambda}
\end{equation}
where
\begin{eqnarray}
\beta &=& a^2(1 - E^2)\;,
\label{eq:beta}\\
\beta z_+ &=& \frac{1}{2}\biggl[L_z^2 + Q + \beta +\sqrt{\left[L_z^2 + Q + \beta\right]^2 - 4\beta Q}\biggr]\;.
\nonumber\\
\label{eq:betazedplus}
\end{eqnarray}
The parameter $\chi$ tends to be a more convenient angle to integrate than $\theta$, since it grows monotonically and has no turning points.

\begin{figure}
\includegraphics[width=0.48\textwidth]{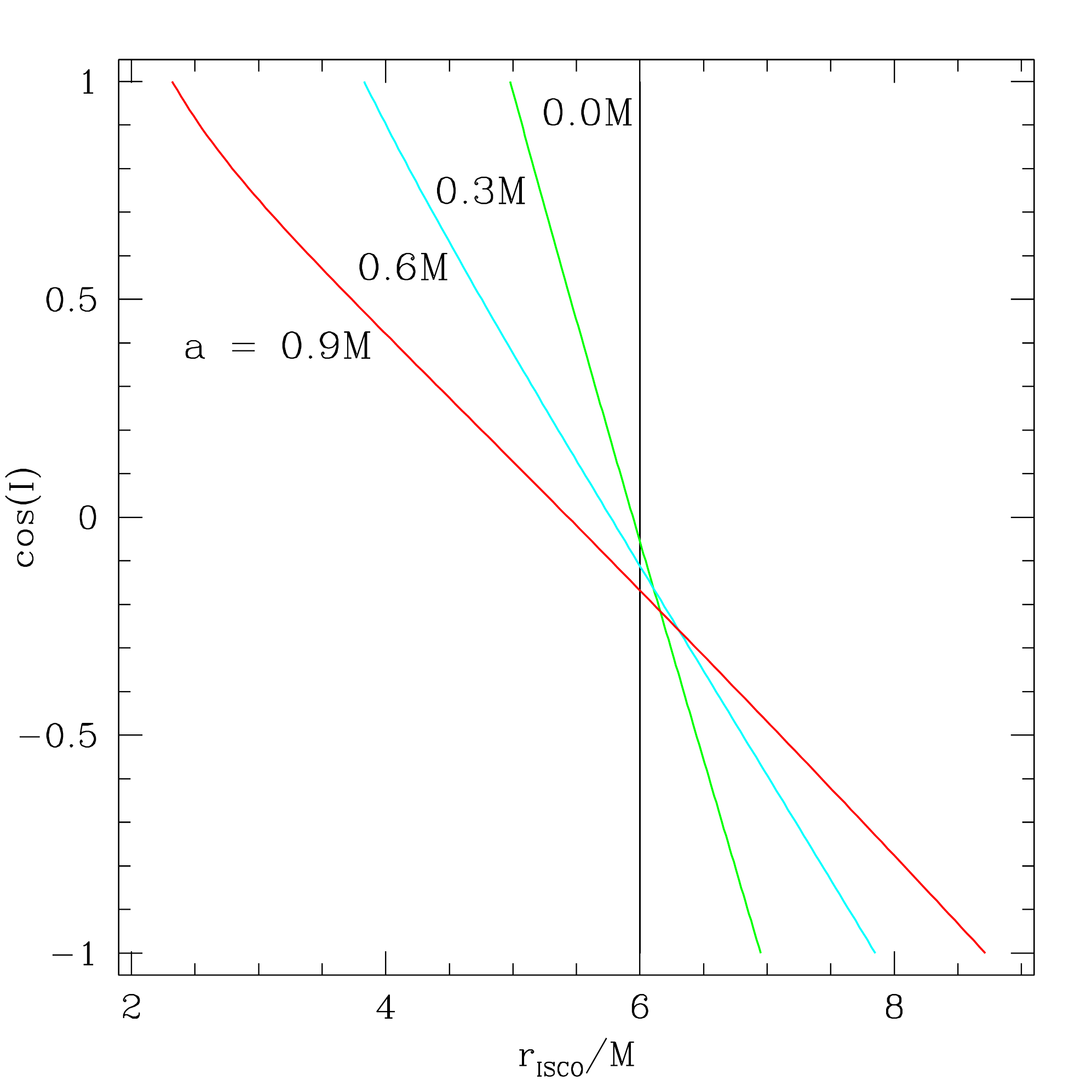}
\caption{The radius of the innermost stable circular orbit, $r_{\rm ISCO}$, as a function of inclination angle $I$ for several values of black hole spin $a$.  Circular orbits are stable for $r_{\rm o} > r_{\rm ISCO}$, and unstable for $r_{\rm o} < r_{\rm ISCO}$.  As described in Secs.\ {\ref{sec:OT}} and {\ref{sec:GOT}}, the transition from slowly evolving stable circular orbits to a plunging geodesic occurs near $r_{\rm ISCO}$.}
\label{fig:separatrix}
\end{figure}

Circular orbits are stable if $R''(r_{\rm o}) < 0$.  The separatrix dividing stable from unstable is the set of orbits which satisfy $R = 0$, $R' = 0$, $R'' = 0$; the orbit radius which satisfies these equations is known as the {\it innermost stable circular orbit}, or ISCO.  The radius of the ISCO is simple to compute for equatorial orbits {\cite{bpt72}}:
\begin{eqnarray}
r^{\rm eq}_{\rm ISCO}/M &=& 3 + Z_2 \mp \left[(3 - Z_1)(3 + Z_1 + 2Z_2)\right]^{1/2}\;,\nonumber\\
\label{eq:risco}
\\
Z_1 &\equiv& 1 + (1 - q^2)^{1/3}\left[(1 + q)^{1/3} + (1 - q)^{1/3}\right]\;,\nonumber\\
\label{eq:Z1}
\\
Z_2 &\equiv& \left(3q^2 + Z_1^2\right)^{1/2}\;.
\label{eq:Z2}
\end{eqnarray}
It is straightforward to solve the system of equations $R = 0$, $R' = 0$, $R'' = 0$ to construct $r_{\rm ISCO}(I)$ for all inclinations.  Figure {\ref{fig:separatrix}} shows the ISCO as a function of $\cos I$ for several black hole spins.

Finally, it is worth noting that circular orbits have a well-defined frequency spectrum, executing axial motion with frequency $\Omega_\phi$, and oscillating in polar angle with frequency $\Omega_\theta$.  See Ref.\ {\cite{fh09}} for explicit formulas for computing these frequencies.  In the limit of equatorial orbits, the frequency $\Omega_\theta$ becomes unimportant since the motion does not exhibit polar oscillations, and
\begin{equation}
\Omega_\phi \to \Omega^{\rm eq}_\phi = \pm\frac{M^{1/2}}{r_{\rm o}^{3/2} \pm aM^{1/2}}\;.
\label{eq:KerrEqOmPhi}
\end{equation}
The discrete frequency spectrum of circular orbits plays an important role in computing GWs from these orbits: the radiation can be expanded in Fourier modes, with contributions from harmonics of the orbit's distinct frequencies.  This means that, for example, the rate at which energy is carried away by GWs can be written
\begin{equation}
\dot E = \sum_{\ell mk} \dot E_{\ell mk}\;,
\end{equation}
where the indices $m$ and $k$ label harmonics of the axial and polar frequencies; the index $\ell$ is a spheroidal harmonic index.  Similar formulas describe the rates at which $L_z$ and $Q$ evolve due to GW emission.  Input from a code which computes such ``fluxes'' of $E$, $L_z$, and $Q$ {\cite{dh06}} plays an important role in our construction of the worldline followed by a body plunging into a Kerr black hole.

\subsection{Slowly evolving circular orbits}
\label{sec:slowevolve}

We wish to consider a large mass-ratio binary whose orbit is accurately described as an inclined, circular Kerr geodesic orbit on ``short'' timescales, but that evolves from one circular orbit to another on ``long'' timescales.  The short timescale will be of order the small body's orbital period, and the long timescale will be of order the radiation reaction timescale.  The ratio of these timescales, whether expressed in terms of coordinate time, proper time, or Mino time, is the mass ratio $\eta$:
\begin{equation}
\frac{T_{\rm o}}{T_{\rm RR}} \sim \eta\;.
\end{equation}
We now examine some consequences of this separation of timescales, and some properties of the smaller body's orbital motion while it is accurately described as slowly evolving from geodesic to geodesic.

Write the orbital radius as a function of Mino time: $r_{\rm o} = r_{\rm o}(\lambda)$.  The orbital radius is governed by the radial geodesic equation:
\begin{equation}
\left(\frac{dr_{\rm o}}{d\lambda}\right)^2 = R(r_{\rm o})\;,
\label{eq:radgeod}
\end{equation}
where $R(r)$ is defined by Eq.\ (\ref{eq:rdot}).  The function $R(r)$ also depends on the orbit's $E$, $L_z$, and $Q$.  Suppose that $E$, $L_z$, and $Q$ themselves vary with $\lambda$, and apply $d/d\lambda$ to both sides of Eq.\ (\ref{eq:radgeod}):
\begin{eqnarray}
2\left(\frac{dr_{\rm o}}{d\lambda}\right)&&\left(\frac{d^2r_{\rm o}}{d\lambda^2}\right) =
\frac{\partial  R}{\partial r}\left(\frac{dr_{\rm o}}{d\lambda}\right)
\nonumber\\
&& + \frac{\partial  R}{\partial E}\left(\frac{dE}{d\lambda}\right) + \frac{\partial  R}{\partial L_z}\left(\frac{dL_z}{d\lambda}\right) + \frac{\partial  R}{\partial Q}\left(\frac{dQ}{d\lambda}\right)\;.
\nonumber\\
\label{eq:radgeod2}
\end{eqnarray}
Each of the terms on the right-hand side are to be evaluated at $r = r_{\rm o}$.  Let us now examine how these terms scale with the mass ratio $\eta$:

\begin{itemize}

\item Assuming that they evolve due to GW emission, the rates of change $dE/d\lambda$, $dL_z/d\lambda$, and $dQ/d\lambda$ are each proportional to $\eta$.

\item The derivatives $\partial R/\partial E$, $\partial R/\partial L_z$, and $\partial R/\partial Q$ are each independent of $\eta$.

\item Because we are considering circular Kerr orbits, the derivative $\partial R/\partial r$ vanishes at $r = r_{\rm o}$.  However, corrections proportional to $\eta$ appear when one accounts for the slow evolution of $E$, $L_z$, and $Q$, so this term is of order $\eta$.

\item The Mino-time radial velocity $dr_{\rm o}/d\lambda$ and acceleration $d^2r_{\rm o}/d\lambda^2$ must each be proportional to $\eta$.

\end{itemize}

Taking these scalings into account shows that some terms in Eq.\ (\ref{eq:radgeod2}) are $O(\eta)$, and others are $O(\eta^2)$.  The equation must hold at each order in $\eta$, so we may separate these different scalings.  Doing so yields the following conditions that our slowly evolving geodesic must satisfy:
\begin{eqnarray}
O(\eta):&&\quad  \frac{\partial  R}{\partial E}\left(\frac{dE}{d\lambda}\right) + \frac{\partial  R}{\partial L_z}\left(\frac{dL_z}{d\lambda}\right) + \frac{\partial  R}{\partial Q}\left(\frac{dQ}{d\lambda}\right) = 0\;,
\nonumber\\
\label{eq:circconstraint}\\
O(\eta^2):&&\quad 
2\left(\frac{dr_{\rm o}}{d\lambda}\right)\left(\frac{d^2r_{\rm o}}{d\lambda^2}\right) =
\frac{\partial  R}{\partial r}\left(\frac{dr_{\rm o}}{d\lambda}\right)\;.
\end{eqnarray}
The $O(\eta)$ constraint (\ref{eq:circconstraint}) is a rule that governs how the constants of the motion evolve for slowly evolving circular orbits.  The $O(\eta^2)$ equation can be rewritten by canceling a common factor on both sides:
\begin{equation}
\frac{d^2r_{\rm o}}{d\lambda^2} = \frac{1}{2} \frac{\partial  R}{\partial r}\;.
\label{eq:circaccel}
\end{equation}
As we will see, Eq.\ (\ref{eq:circaccel}) is a very useful tool for characterizing the evolution of circular orbits when their properties evolve due to GW emission.

\section{Equations of motion for the transition from inspiral and the final plunge}
\label{sec:transitionEOM}

When the system can no longer be regarded as slowly evolving through a sequence of geodesics, what is the correct way to model its motion?  In this section, we first summarize the analysis of Ori and Thorne.  We then show how to generalize their analysis for circular orbits of arbitrary inclination, and conclude by discussing the system's final plunge dynamics.

\subsection{The Ori-Thorne calculation}
\label{sec:OT}

Following OT00, rewrite Eq.\ (\ref{eq:rdot}) for equatorial orbits ($\theta = \pi/2$) as
\begin{equation}
\left(\frac{dr_{\rm o}}{d\tau}\right)^2 = E^2 - V_r(r_{\rm o}, E, L_z)\;.
\label{eq:rdot_OT00}
\end{equation}
The potential $V_r$ is simply related to the function $R$:
\begin{equation}
V_r = E^2 - \frac{R(r_{\rm o})}{r_{\rm o}^4}\;.
\label{eq:radial1}
\end{equation}
Circular orbits exist where $V_r = E^2$ and $\partial V_r/\partial r = 0$.  Applying $d/d\tau$ to both sides of Eq.\ (\ref{eq:rdot_OT00}) and repeating the analysis of Sec.\ {\ref{sec:slowevolve}} but using the potential $V_r$, we find the following equation governs the radial acceleration of the small body:
\begin{equation}
\frac{d^2r_{\rm o}}{d\tau^2} = -\frac{1}{2}\frac{\partial V_r}{\partial r}\;.
\label{eq:radial2}
\end{equation}
The right-hand side is evaluated at $r = r_{\rm o}$, and thus vanishes at leading order in $\eta$.  As we will see below, corrections enter when we introduce the evolution of $E$ and $L_z$ due to GW emission.  We also find an equation analogous to (\ref{eq:circconstraint}):
\begin{equation}
\frac{\partial V_r}{\partial E}\left(\frac{dE}{d\tau}\right) + \frac{\partial V_r}{\partial L_z}\left(\frac{dL_z}{d\tau}\right) = 0\;.
\label{eq:circconstraint2}
\end{equation}

Consider a system that is approaching the ISCO via a sequence of circular orbits.  We put
\begin{eqnarray}
r_{\rm o} &=& r_{\rm ISCO} + x\;,
\label{eq:rnearISCO}
\\
L_z &=& L_z^{\rm ISCO} + \delta L_z\;,
\label{eq:LznearISCO}
\\
E &=& E^{\rm ISCO} + \Omega^{\rm ISCO}_\phi\delta L_z\;.
\label{eq:EnearISCO}
\end{eqnarray}
Here, $\Omega^{\rm ISCO}_\phi$ is the frequency $\Omega_\phi$ given in Eq.\ (\ref{eq:KerrEqOmPhi}) evaluated at $r_{\rm ISCO}$.  We have used the fact that for circular and equatorial orbits
\begin{equation}
\left(\frac{dE}{dt}\right)^{\rm GW} = \Omega_\phi \left(\frac{dL_z}{dt}\right)^{\rm GW}\;,
\label{eq:circfluxes}
\end{equation}
and we assume that GW backreaction is the only mechanism by which $E$ and $L_z$ evolve.

As we enter the transition, we assume that Eq.\ (\ref{eq:radial2}) continues to govern the behavior of the orbit.  Our goal will be to evaluate $\partial V_r/\partial r$ with $r_{\rm o}$, $L_z$, and $E$ given by Eqs.\ (\ref{eq:rnearISCO})--(\ref{eq:EnearISCO}).  Begin by examining the behavior of $V_r(r_{\rm o},E,L_z)$ near the ISCO.  Use
\begin{eqnarray}
V_r|_{\rm ISCO} &=& (E^{\rm ISCO})^2\;,
\nonumber\\
\frac{\partial V_r}{\partial r}\Biggl|_{\rm ISCO} &=& 0\;,
\nonumber\\
\frac{\partial^2 V_r}{\partial r^2}\Biggl|_{\rm ISCO} &=& 0\;,
\end{eqnarray}
where the notation ``$|_{\rm ISCO}$'' means that these terms are evaluated by setting all of $r_{\rm o}$, $E$, and $L_z$ to their ISCO values.  Expanding $V_r$ in $r_{\rm o}$, $E$, and $L_z$, we find
\begin{eqnarray}
V_r &=& \left(E^{\rm ISCO}\right)^2 + \frac{1}{6}\left(\frac{\partial^3 V_r}{\partial r^3}\right) x^3
\nonumber\\
& & + \left(\frac{\partial^2V_r}{\partial L_z\partial r} + \Omega_\phi \frac{\partial^2V_r}{\partial E\partial r}\right) x\,\delta L_z\nonumber\\
& & +\, \mbox{Terms that do not depend on $x$}\;.
\label{eq:VrnearISCO}
\end{eqnarray}
All terms in parentheses in Eq.\ (\ref{eq:VrnearISCO}) are to be evaluated at the ISCO.  A term linear in $\delta L_z$ but independent of $x$ vanishes thanks to Eq.\ (\ref{eq:circconstraint2}).

Next, substitute Eq.\ (\ref{eq:VrnearISCO}) into Eq.\ (\ref{eq:radial2}), noting that $d^2r_{\rm o}/d\tau^2 = d^2x/d\tau^2$ and $\partial/\partial r = \partial/\partial x$:
\begin{equation}
\frac{d^2x}{d\tau^2} = -\frac{1}{4} \left(\frac{\partial^3  V_r}{\partial r^3}\right) x^2 - \frac{1}{2} \left(\frac{\partial^2V_r}{\partial L_z\partial r} + \Omega_\phi \frac{\partial^2V_r}{\partial E\partial r}\right)\delta L_z\;.
\label{eq:radial3}
\end{equation}
As in the preceding equation, terms in parentheses here are to be evaluated at the ISCO.

To manipulate further, we need to model $\delta L_z$, the angular momentum lost by the orbit as the small body evolves through the transition.  This regime is, by definition, short-lived, so the angular momentum flux is nearly equal to its value on the ISCO throughout the transition.  Following OT00, we write
\begin{eqnarray}
\delta L_z &=& \eta \left(\frac{dL_z}{d\tau}\right)^{\rm ISCO}(\tau -
\tau_{\rm ISCO})
\nonumber\\
&=& \eta \left(\frac{dL_z}{dt}\right)^{\rm ISCO}
\left(\frac{dt}{d\tau}\right)^{\rm ISCO}(\tau - \tau_{\rm ISCO})
\nonumber\\
&=& \eta \kappa (\tau - \tau_{\rm ISCO})\;.
\label{eq:Lz_OT}
\end{eqnarray}
Here, $\tau_{\rm ISCO}$ gives the value of the body's proper time when its angular momentum is equal to the value that a geodesic orbit would have at the ISCO.  The quantities $(dL_z/d\tau)^{\rm ISCO}$ and $(dL_z/dt)^{\rm ISCO}$ describe the flux of angular momentum in GWs.  The second line is written in a more useful form for our purposes since our radiation emission code computes this flux per unit of coordinate time $t$.  For later convenience, we have scaled out a factor of the mass ratio $\eta$ which normalizes the GW flux.  On the last line, we have defined
\begin{equation}
\kappa = \left(\frac{dL_z}{dt}\right)^{\rm ISCO} \left(\frac{dt}{d\tau}\right)^{\rm ISCO}\;.
\end{equation}

Combining all of these pieces, we find
\begin{equation}
\frac{d^2x}{d\tau^2} = -\alpha_{\rm OT} x^2 - \eta \beta_{\rm OT} \kappa\left(\tau - \tau_{\rm ISCO}\right)\;,
\label{eq:radial4}
\end{equation}
where
\begin{eqnarray}
\alpha_{\rm OT} &=& \frac{1}{4}\left(\frac{\partial^3 V_r}{\partial r^3}\right)\;,
\label{eq:alphadef}\\
\beta_{\rm OT} &=& \frac{1}{2}\left(\frac{\partial^2V_r}{\partial L_z\partial r} + \Omega_\phi \frac{\partial^2V_r}{\partial E\partial r}\right)\;,
\label{eq:betadef}
\end{eqnarray}
and where all quantities in parentheses are to be evaluated at the ISCO.

Let us now rescale the radial variable $x$ and the proper time $\tau$ as follows:
\begin{eqnarray}
x/M &\equiv& \eta^{2/5}\,(\beta_{\rm OT}\kappa)^{2/5}\alpha^{-3/5}\,X
\label{eq:Xdef}\\
\left(\tau - \tau_{\rm ISCO}\right)/M &\equiv& \eta^{-1/5}\,(\alpha_{\rm OT}\beta_{\rm OT}\kappa)^{-1/5} T\;.
\label{eq:Tdef}
\end{eqnarray}
These definitions bring Eq.\ (\ref{eq:radial4}) into the form
\begin{equation}
\frac{d^2X}{dT^2} = -X^2 - T\;.
\label{eq:ottrans}
\end{equation}
This is the main result of OT00, their equation (3.22).

\begin{figure}
\includegraphics[width=0.48\textwidth]{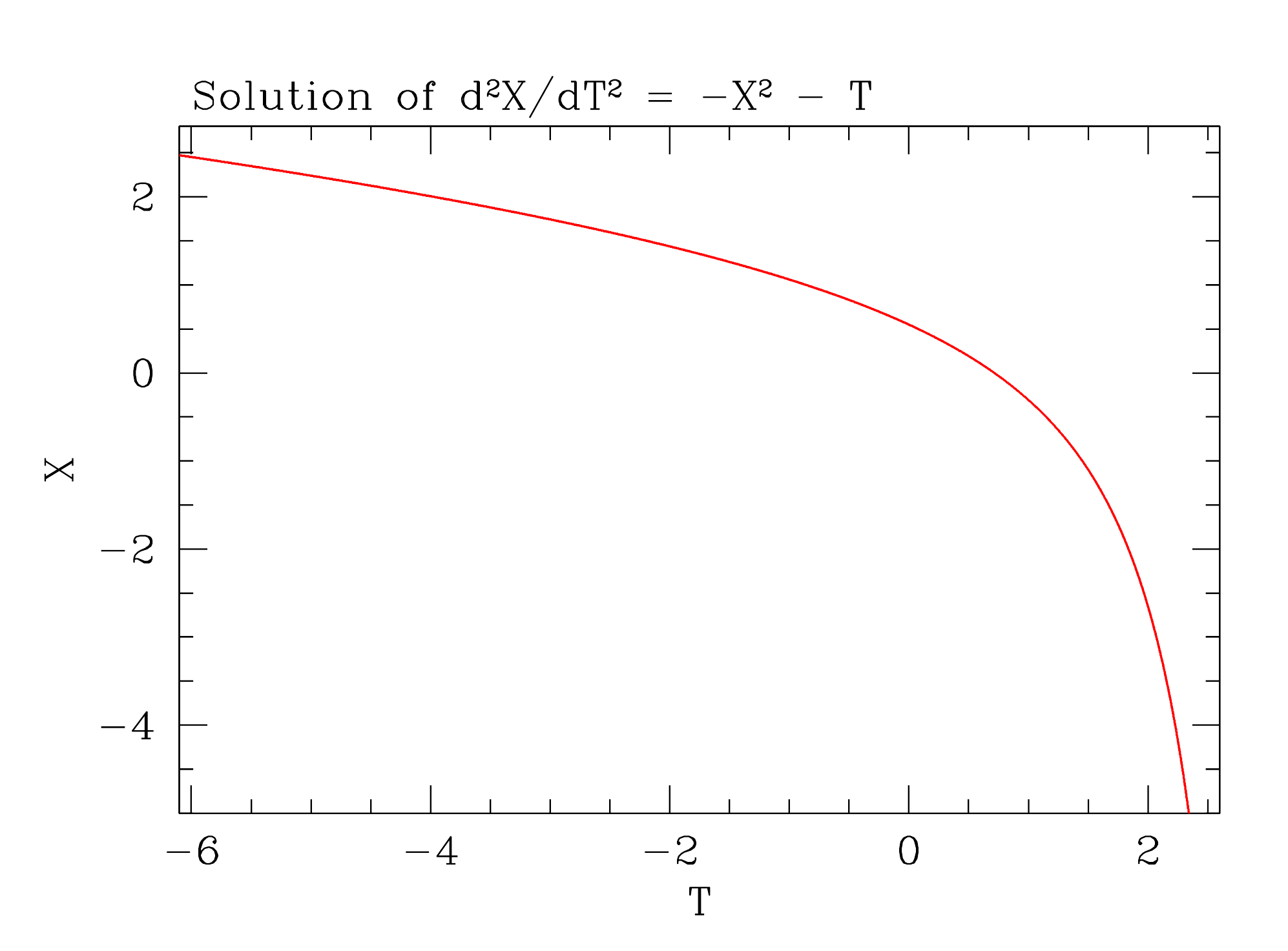}
\caption{The function $X(T)$ found by numerically solving Eq.\ (\ref{eq:ottrans}).  Note that this solution only needs to be constructed once: with the scalings defined in Eqs.\ (\ref{eq:Xdef}) and (\ref{eq:Tdef}), this solution describes the transition for all values of $a$ and $\eta$ (provided the large mass-ratio approximation remains valid).}
\label{fig:XvsT}
\end{figure}

To solve this equation, we use the fact that large negative $T$ corresponds to the beginning of the transition regime.  OT00 note that at this moment (when the system is still evolving adiabatically), the orbit sits at the minimum of $V_r$, and the system's motion has the solution
\begin{equation}
X = \sqrt{-T}\;.
\label{eq:Xadiabatic}
\end{equation}
By iteration, one can find a solution which more accurately solves Eq.\ (\ref{eq:ottrans}):
\begin{eqnarray}
X &=& \sqrt{-T} + \frac{1}{8T^2} - \frac{49}{128(-T)^{9/2}} - \frac{1225}{256T^7}
\nonumber\\
& & - \frac{4412401}{32768(-T)^{19/2}} + \ldots
\label{eq:startX}
\end{eqnarray}
Using Eq.\ (\ref{eq:startX}) to find $X$ and $dX/dT$ at some inital time $T_{\rm i}$, it is simple to solve Eq.\ (\ref{eq:ottrans}) numerically.  Figure {\ref{fig:XvsT}} shows the resulting $X(T)$ through the transition regime.

\subsection{Generalizing Ori-Thorne}
\label{sec:GOT}

Ori and Thorne's analysis must be generalized in order to study the transition and plunge of inclined orbits.  Such a generalization was first attempted by Sundararajan {\cite{pranesh}}.  Unfortunately, we have found that this generalization is flawed: the point at which transition begins using the algorithm of Ref.\ {\cite{pranesh}} depends strongly on the initial value of the polar angle $\theta$.  This means that two trajectories with the same initial values of $E$, $L_z$, and $Q$ but different initial values of $\theta$ will undergo different plunge dynamics.

This behavior arises because Sundararajan's equations of motion do not separate the radial and angular degrees of freedom.  His prescription mixes orbit terms which vary on dynamical orbit timescale $T_{\rm orb} \sim M$ with terms that vary on a dissipative GW timescale $T_{\rm GW} \sim M/\eta$.  Although in principle the short-timescale orbital dynamics may have an impact on the transition, one needs to include {\it all}\ short-timescale terms in order to properly model these effects.  For example, the self force includes both dissipative elements that vary only on the timescale $T_{\rm GW}$ as well as oscillatory terms that vary on the timescale $T_{\rm orb}$.  Sundararajan's analysis thus has both short- and long-timescale orbit effects, but only includes the long-timescale dissipative effects.

In our analysis, we model dissipation using only the orbit-averaged impact of radiation reaction.  Since our model for the dissipative evolution averages out short-timescale effects, consistency mandates that we develop a way to describe the transition that decouples the orbit's long-timescale dissipative radial motion from the orbit's other degrees of freedom.  We do so using the Mino-time description given in Sec.\ {\ref{sec:kerr}}.  Begin with the acceleration equation derived in Sec.\ {\ref{sec:slowevolve}},
\begin{equation}
\frac{d^2r_{\rm o}}{d\lambda^2} = \frac{1}{2}\frac{\partial R}{\partial r}\;.
\label{eq:minoradial}
\end{equation}
Exactly as in OT00, we now expand all relevant quantities near the ISCO:
\begin{eqnarray}
r_{\rm o} &=& r_{\rm ISCO} + x\;,
\label{eq:inclrisco}
\\
E &=& E^{\rm ISCO} + \delta E\;,
\label{eq:inclEisco}
\\
L_z &=& L_z^{\rm ISCO} + \delta L_z\;,
\label{eq:inclLzisco}
\\
Q &=& Q^{\rm ISCO} + \delta Q\;.
\label{eq:inclQisco}
\end{eqnarray}
The quantities $\delta E$, $\delta L_z$, and $\delta Q$ describe how the energy, axial angular momentum, and Carter constant evolve through the transition.  In the equatorial case, $\delta E$ and $\delta L_z$ are connected by Eq.\ (\ref{eq:circfluxes}).  A similar connection exists between $(\delta E, \delta L_z, \delta Q)$ for inclined circular orbits, but is inclination dependent and not simple to write down in closed form, although it is worth noting that $(\delta E, \delta L_z, \delta Q)$ are constrained by Eq.\ (\ref{eq:circconstraint}).  These connections are built into the code that we use to compute these quantities {\cite{dh06}}.

Let us now expand $R$ around the ISCO.  Circular orbits at the ISCO are defined by
\begin{eqnarray}
R = 0\;,\qquad R' = 0\;,\qquad R'' = 0\;.
\label{eq:circincl}
\end{eqnarray}
Taking advantage of these conditions and using Eqs.\ (\ref{eq:circconstraint}) and (\ref{eq:inclrisco})--(\ref{eq:inclQisco}), we find
\begin{eqnarray}
R &=& \frac{1}{6}\left(\frac{\partial^3 R}{\partial r^3}\right) x^3
\nonumber\\
&+&
\left(\frac{\partial^2R}{\partial r\partial E}\delta E +
\frac{\partial^2R}{\partial r\partial L_z}\delta L_z +
\frac{\partial^2R}{\partial r\partial Q}\delta Q\right)x
\nonumber\\
& + & \mbox{Terms that do not depend on $x$}
\label{eq:Rexpand}
\end{eqnarray}
near the ISCO.  All terms in parentheses are evaluated on the ISCO.  Put
\begin{eqnarray}
\delta E &=& \eta \left(\frac{dE}{d\lambda}\right)^{\rm ISCO}(\lambda - \lambda_{\rm ISCO})\;,
\label{eq:E_GOT}
\\
\delta L_z &=& \eta \left(\frac{dL_z}{d\lambda}\right)^{\rm ISCO}(\lambda - \lambda_{\rm ISCO})\;,
\label{eq:Lz_GOT}
\\
\delta Q &=& \eta \left(\frac{dQ}{d\lambda}\right)^{\rm ISCO}(\lambda - \lambda_{\rm ISCO})\;.
\label{eq:Q_GOT}
\end{eqnarray}
Notice that we have written the rates of change as $d/d\lambda$, rather than $d/dt$.  For convenience, let us write
\begin{equation}
\delta {\cal C} = \eta\kappa_{\cal C}(\lambda - \lambda_{\rm ISCO})
\end{equation}
for ${\cal C} \in [E,L_z,Q]$.  The value of $\kappa_{\cal C}$ can be easily read out for each ${\cal C}$.  Using $d^2r_{\rm o}/d\lambda^2 = d^2x/d\lambda^2$ plus the fact that $\partial/\partial r = \partial/\partial x$ near the ISCO, then combining Eqs.\ (\ref{eq:minoradial}) and (\ref{eq:Rexpand}) we find
\begin{equation}
\frac{d^2x}{d\lambda^2} = -A x^2 - \eta B(\lambda - \lambda_{\rm ISCO})\;,
\end{equation}
where
\begin{eqnarray}
A &\equiv& -\frac{1}{4}\left(\frac{\partial^3 R}{\partial r^3}\right)\;,
\\
B &\equiv& -\frac{1}{2}\left(\frac{\partial^2 R}{\partial r\partial E}\kappa_E + \frac{\partial^2 R}{\partial r\partial L_z}\kappa_{L_z} + \frac{\partial^2 R}{\partial r\partial Q}\kappa_Q\right)\;.
\nonumber\\
\end{eqnarray}
All terms in parentheses are evaluated at the ISCO.  We have chosen signs in these definitions to insure that $A$ and $B$ are both positive.

Next, scale the radial variable $x$ and Mino-time $\lambda$ as follows:
\begin{eqnarray}
x/M &\equiv& \eta^{2/5} B^{2/5}A^{-3/5} X\;,
\label{eq:XdefGOT}
\\
M(\lambda - \lambda_{\rm ISCO}) &\equiv& \eta^{-1/5}(AB)^{-1/5} L\;.
\label{eq:LdefGOT}
\end{eqnarray}
The equation governing the transition becomes
\begin{equation}
\frac{d^2X}{dL^2} = -X^2 - L\;.
\label{eq:ottransgen}
\end{equation}
This is identical in form to Eq.\ (\ref{eq:ottrans}) and thus admits identical solutions.  The only difference is in the definition of the timelike coordinate: the variable $T$ used in Eq.\ (\ref{eq:ottrans}) is a rescaled proper time $\tau$, and the variable $L$ used here is a rescaled Mino time $\lambda$.  As a consequence, the solution for $X$ shown in Fig.\ {\ref{fig:XvsT}} carries over to inclined orbits with no modification other than relabeling the independent parameter.  The only change needed is to use the scalings (\ref{eq:XdefGOT}) and (\ref{eq:LdefGOT}) to describe the motion in the physical space near the black hole.

\subsection{From transition to plunge}
\label{sec:plunge}

Equation (\ref{eq:ottrans}) was derived by Taylor expanding the function $R$ in the vicinity of the ISCO.  As such, as $x$ gets large, neglected terms on the right-hand side of (\ref{eq:ottrans}) are expected to become important.  In addition, the impact of radiative backreaction should decrease in importance as the small body comes closer to the horizon.  Equations (\ref{eq:Lz_OT}), (\ref{eq:E_GOT}), (\ref{eq:Lz_GOT}), and (\ref{eq:Q_GOT}) will then no longer accurately model how $E$, $L_z$, and $Q$ behave.  It may be more accurate to treat these quantities as constants.  The small body's motion would become a free-fall geodesic plunging into the black hole.  The radial motion would then be governed by Eq.\ (\ref{eq:minoradial}), with ($E, L_z, Q$) and $dr/d\lambda$ chosen to have their values at the moment when the transition description becomes inaccurate.

This suggests that it would make sense to switch the motion from that described by Eq.\ (\ref{eq:ottrans}) to a plunging geodesic when the transition has proceeded ``far enough."  When we are confident that Eq.\ (\ref{eq:ottrans}) is becoming inaccurate, we freeze the small body's values of $E$, $L_z$, and $Q$, and treat the motion as a simple geodesic which plunges into the larger black hole.  In Sec.\ {\ref{sec:WLstep1}}, we discuss how to define what ``far enough'' means, and thus how to choose when the transition ends and plunge begins.

\section{Making the full inspiral and plunge worldline}
\label{sec:worldline}

We now synthesize all the elements described above and describe how we construct the worldline of a body that inspirals and then plunges into a Kerr black hole.  The procedure we follow has two major steps.  We first compute the radial motion of the small body as described in Sec.\ {\ref{sec:transitionEOM}}.  The output of this step is a description of the body's motion parameterized using Mino time: $[r_{\rm o}(\lambda), E(\lambda), L_z(\lambda), Q(\lambda)]$.  We then convert from Mino time $\lambda$ to Boyer-Lindquist coordinate time $t$, putting our description of the small body's motion into the form needed for the BHPT code that we use to study black hole mode excitation.  At the same time, we also compute the small body's angular motion.  The final product of this procedure is a full description of the coordinate motion as a function of time $t$, $[r_{\rm o}(t), \theta(t), \phi(t)]$.

\subsection{Radial motion in Mino time}
\label{sec:WLstep1}

The Mino-time radial motion is computed by stitching together solutions describing the inspiral, transition, and plunge motions.

\subsubsection{Adiabatic inspiral}

We begin by computing the trajectory $[r_{\rm o}(\lambda), E(\lambda), L_z(\lambda), Q(\lambda)]$ that the small body would follow if the inspiral accurately described the motion from large radius all the way down to the ISCO.  To do this, we use a frequency-domain BHPT code {\cite{dh06}} to compute the rate at which GWs evolve the orbit's energy, angular momentum, and Carter constant.  We choose an initial orbit, $[r_{{\rm o},0}, E_0, L_{z,0}, Q_0]$, integrate using the rates of change\footnote{Because we restrict ourselves to quasi-circular orbits, we strictly need only $dE/d\lambda$ and $dL_z/d\lambda$.  The requirement that a circular orbit must evolve into another circular orbit under adiabatic radiation reaction \cite{ko96, ryan96} then yields $dr/d\lambda$ and $dQ/d\lambda$ \cite{h2000}.} $[dr/d\lambda, dE/d\lambda, dL_z/d\lambda, dQ/d\lambda]$ predicted by BHPT, and thereby produce ``inspiral-only" worldlines describing the small body's radial motion from large radius down to the ISCO.

A useful property of this ``inspiral-only'' worldline is that, once computed, it is simple to rescale in order to describe any mass ratio (within the confines of the large mass-ratio limit needed for BHPT to be valid).  Each of the rates of change $[dE/d\lambda, dL_z/d\lambda, dQ/d\lambda]$ is proportional to the mass ratio $\eta$.  We can thus define a renormalized Mino-time interval $d\lambda' = \eta^{-1}d\lambda$.  Once we have computed the worldline using the mass-ratio-scaled Mino time, $[r(\lambda'), E(\lambda'), L_z(\lambda'), Q(\lambda')]$, it is straightforward to rescale the time axis in order to describe the adiabatic inspiral for any $\eta$.

\subsubsection{Transition}

The inspiral-only worldline will accurately describe the small body's motion until it comes close to the ISCO, and the orbit no longer adiabatically follows the evolving extremum of the radial function $R$.  Once this occurs, we switch to the transition solution described in Sec.\ {\ref{sec:GOT}}.

We need to choose a value $L_i$ to designate when inspiral ends and the transition begins.  Our choice is driven by two competing factors:

\begin{itemize}

\item If $|L_i|$ is too large, then our transition-domain description of the fluxes (\ref{eq:E_GOT}) -- (\ref{eq:Q_GOT}) does not match the fluxes that we compute from black hole perturbation theory.  We have found that if $L_i > -4$, then the mismatch between the fluxes does not exceed $5\%$.

\item If $|L_i|$ is too small, then we do not begin the transition until after the adiabatic approximation to the inspiral has begun to break down.  The corrections beyond $\sqrt{-L}$ that appear in Eq.\ (\ref{eq:startX}) (updating from the time-like variable $T$ to $L$) quantify the importance of post-adiabatic corrections to the small body's motion.  When $L_i = -1.4$, the first correction to the solution in this equation is $5\%$ as large as the leading term.

\end{itemize}

Choosing $5\%$ mismatches and errors (an admittedly arbitrary choice) suggests choosing the initial time for the transition in the range $-4 \le L_i -1.4$.  In Sec.\ {\ref{sec:results}}, we more carefully investigate how our worldlines vary over the range $-5 \le L_i \le -1$.

It is likewise important to choose a value $L_f$ to define when the transition ends and the plunge begins.  We again make a choice which is driven by two factors:

\begin{itemize}

\item OT00 notes that, in the dimensionless equation of motion, free fall corresponds to neglecting the final term on the right-hand side of Eq.\ (\ref{eq:ottrans}) [or equivalently, the final term on the right-hand side of Eq.\ (\ref{eq:ottransgen})].  One can thus assess how close one is to free fall by comparing the magnitude of the two terms on the right-hand side of this equation.  We find that by requiring $L_f \ge 2.2$, the correction is never larger than $5\%$ of the free-fall term; it rapidly decreases if we choose a larger value of $L_f$.

\item For large $L_f$, higher-order terms in the Taylor expansion which we neglected in deriving Eq.\ (\ref{eq:ottransgen}) become important.  The leading corrections to the right-hand side of this equation are of the form $\alpha X^3 + \beta XL$ (where $\alpha$ and $\beta$ depend upon the black hole spin and the location of the ISCO).  To keep these neglected terms smaller than $5\%$ of the leading term, we find that we must have $L_f \le 2.5$.

\end{itemize}

Again, based on a fairly arbitrary $5\%$ mismatch criterion, we advocate choosing a final time for the transition in the range $2.2 \le L_f \le 2.5$.  We also investigate carefully how this choice affects our worldlines in Sec.\ {\ref{sec:results}}.

With this in mind, here is the algorithm we use in this paper to compute the Mino-time worldline in the transition regime:

\begin{itemize}

\item Choose a value $L_i$ in the range $-5 \le L_i \le -1$ to designate the end of the inspiral and the beginning of the transition.  Using Eq.\ (\ref{eq:LdefGOT}), compute the corresponding Mino time $\lambda_i$.

\item Using the solution $X(L)$ (shown in Fig.\ {\ref{fig:XvsT}}) which solves the generalized Ori-Thorne transition equation (\ref{eq:ottransgen}), compute the corresponding value of $X_i = X(L_i)$.

\item Choose a value $L_f$ in the range $2.2 \le L_f \le 2.5$ to designate the end of the transition and beginning of the plunge; use Eq.\ (\ref{eq:LdefGOT}) to compute the corresponding Mino time $\lambda_f$.

\item Using Eqs.\ (\ref{eq:XdefGOT}) and (\ref{eq:LdefGOT}), convert $X(L)$ on the range $L_i \le L \le L_f$ into $r(\lambda)$ for the corresponding transition interval.

\end{itemize}

This algorithm is identical to the OT00 procedure for computing $r_{\rm o}$ during the transition, modulo the use of Mino time and the generalization to inclined orbits.  We have found empirically that we must refine the OT00 approach to modeling the evolution of the orbit's integrals energy, angular momentum, and Carter constant during the transition.  In the OT00 model, these quantities are taken to evolve linearly with proper time:
\begin{equation}
E_{\rm OT} = E_{\rm ISCO} + \tau\left(\frac{dE}{d\tau}\right)_{\rm ISCO}\;,
\label{eq:energy_OT}
\end{equation}
where $\left(dE/d\tau\right)_{\rm ISCO}$ is the rate of change of the orbital energy computed by BHPT for an orbit that sits right on the ISCO.  Analogous formulas describe the evolution of the orbit's angular momentum and Carter constant.  Proper time is defined so that $E_{\rm OT} = E_{\rm ISCO}$ when $\tau = 0$.

\begin{figure}
\includegraphics[width=0.48\textwidth]{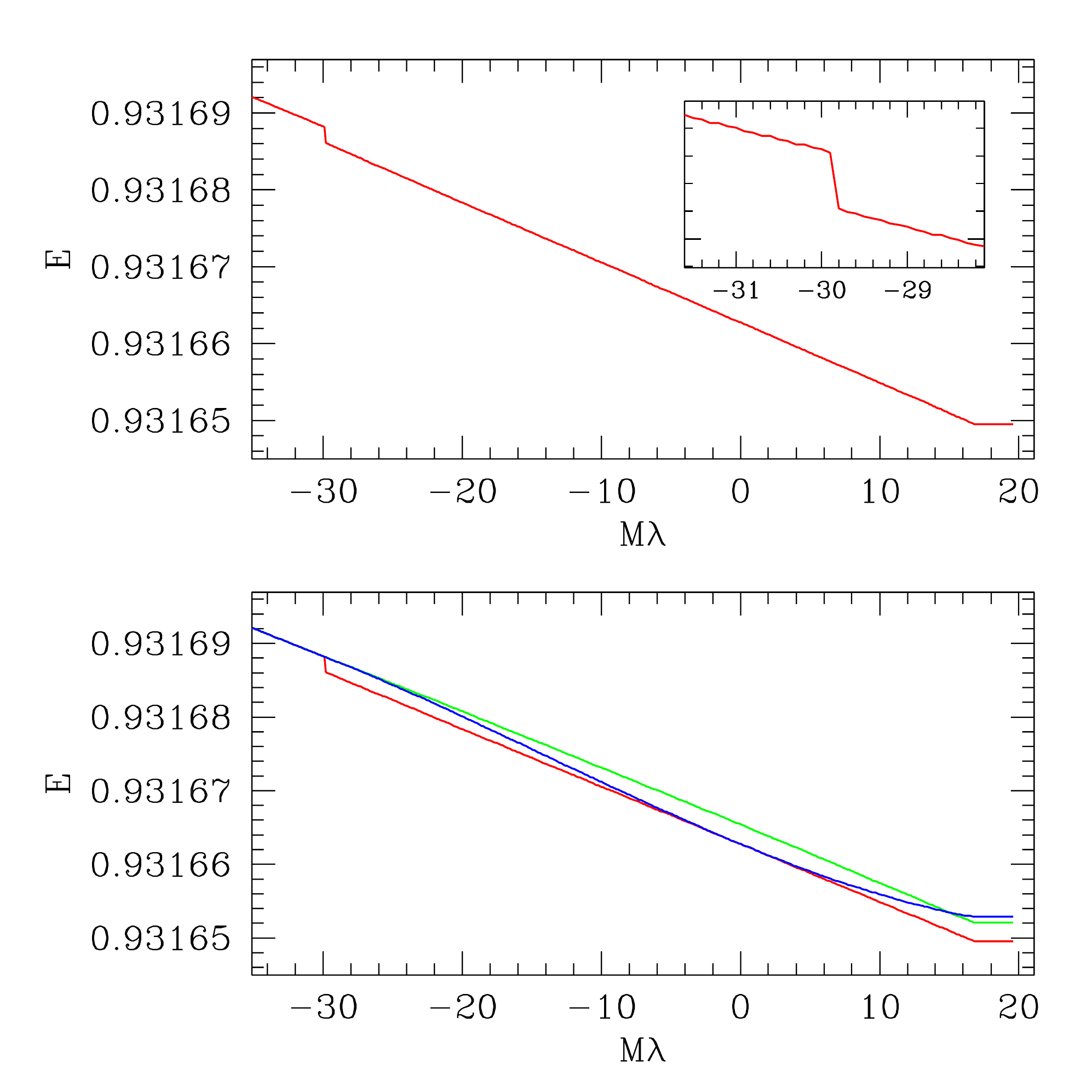}
\caption{An illustration of the three models we consider for evolving orbital energy across the transition.  In all three cases, we consider a binary with mass ratio $\mu/M = 10^{-5}$ and black hole spin $a/M = 0.5$; the system begins at $r_{\rm o} = 5.864M$ with inclination angle $I = 60^\circ$.  In the top panel we show how the system's energy evolves using the original OT00 model, Eq.\ (\ref{eq:energy_OT}).  Notice that $E$ is discontinuous at $\lambda \simeq -29.8/M$, when inspiral ends and the transition region begins.  This discontinuity has observational consequences, since it affects the conversion between Mino time $\lambda$ and Boyer-Lindquist time $t$.  The bottom panel compares this to the models we introduce to enforce continuity of both $E$ and its flux $dE/d\lambda$ when transition begins.  The green curve shows the evolution when we introduce a constant offset and a quadratic correction, Eq.\ (\ref{eq:energy_model1}); the blue curve shows $E$ introducing a quadratic and a cubic correction, Eq.\ (\ref{eq:energy_model2}).}
\label{fig:energy_models}
\end{figure}

We have found that this approach leads to discontinuities in $E$, $L_z$, and $Q$ which noticeably affect the worldlines that we compute: the conversion between Mino time and Boyer-Lindquist time depends upon the values of $E$, $L_z$, and $Q$, so discontinuities in their evolution lead to unphysical artifacts in the coordinate-time-domain worldline\footnote{Note that this discontinuous behavior is part of the original OT00 model.  This was noted by Sundarajan {\cite{pranesh}}, as well as others who have used the OT00 model {\cite{callister}}.}.  An example of this behavior is shown in the top panel of Fig.\ {\ref{fig:energy_models}}.  The red line in this figure shows the evolution of a particular binary's orbital energy $E$; the inspiral ends at $M\lambda \simeq -29.8$, and is marked by a jump in energy as we change from the adiabatic inspiral to the transition.  The discontinuities arise because these orbital integrals do not vary precisely linearly with time; there is some curvature in the time dependence of $E$, $L_z$, and $Q$ as the ISCO is approached.  By our choice of $L_i$, this discontinuity is never more than a $5\%$ effect; however, even this is large enough to have a deleterious impact on the worldlines we construct.

To correct this, we have explored two refinements to the OT00 approach.  Both insure that the integrals and their first derivatives with respect to $\lambda$ are continuous as we move from inspiral to transition.  The first refinement we explore (``Model 1") introduces a constant offset and a quadratic correction in Mino time:
\begin{equation}
E^{\rm M1} = E_{\rm ISCO} + \lambda\left(\frac{dE}{d\lambda}\right)_{\rm ISCO} + C_{E} + \frac{\lambda^2}{2}
{\cal E}^{\rm M1}_2\;;
\label{eq:energy_model1}
\end{equation}
similar forms are used for $L_z$ and $Q$.  In this model, $\lambda = 0$ is the moment when the first derivative of the integrals equals the prediction from BHPT.  The constant ${\cal E}^{\rm M1}_2$ is an estimator for the second derivative of $E$ with $\lambda$ at the end of inspiral.  Recalling that $\lambda_i$ is the value of Mino time corresponding to our chosen  $L_i$, we have
\begin{equation}
{\cal E}^{\rm M1}_2 = \frac{1}{\lambda_i}\left[\left(\frac{dE}{d\lambda}\right)_{\lambda_i} - \left(\frac{dE}{d\lambda}\right)_{\rm ISCO}\right]\;.
\label{eq:E2M1def}
\end{equation}
We similarly define quantities ${\cal L}^{\rm M1}_2$ and ${\cal Q}^{\rm M1}_2$ to smooth the behavior of the angular momentum and Carter constant.  We choose the constant $C_E$ so that the energy is continuous at $\lambda = \lambda_i$:
\begin{equation}
C_E = E(\lambda_i) - E_{\rm ISCO} - \lambda_i\left(\frac{dE}{d\lambda}\right)_{\rm ISCO} - \frac{\lambda_i^2}{2}{\cal E}_2\;.
\label{eq:CEdef}
\end{equation}
We likewise define $C_L$ and $C_Q$ to insure continuity of $L_z$ and $Q$.

Our second refinement (``Model 2'') uses a quadratic and a cubic correction to enforce continuity of $E$ and $dE/d\lambda$ at $\lambda_i$:
\begin{equation}
E^{\rm M2} = E_{\rm ISCO} + \lambda\left(\frac{dE}{d\lambda}\right)_{\rm ISCO} + \frac{\lambda^2}{2}
{\cal E}^{\rm M2}_2 + \frac{\lambda^3}{6}{\cal E}_3\;.
\label{eq:energy_model2}
\end{equation}
Again, we use similar forms for $L_z$ and $Q$.  In this model, when $\lambda = 0$, the first derivative of the integrals equals the prediction from BHPT, and the integrals take their ISCO geodesic values.

The quantities ${\cal E}^{\rm M2}_2$ and ${\cal E}_3$ are estimators for the second and third derivatives of $E$ with $\lambda$, and can be found by equating $E^{\rm M2}(\lambda_i)$ to the orbital energy at the end of inspiral, and equating $(dE^{\rm M2}/d\lambda)_{\lambda_i}$ to the BHPT prediction for the rate of change at the end of inspiral.  Solving the two resulting equations for ${\cal E}^{\rm M2}_2$ and ${\cal E}_3$ yields
\begin{widetext}
\begin{eqnarray}
{\cal E}^{\rm M2}_2 &=& \frac{2}{\lambda_i^2}\left\{3E(\lambda_i) - 3E_{\rm ISCO} - \lambda_i\left[2\left(\frac{dE}{d\lambda}\right)_{\rm ISCO} + \left(\frac{dE}{d\lambda}\right)_{\lambda_i}\right]\right\}\;,
\label{eq:E2M2def}\\
{\cal E}_3 &=& \frac{6}{\lambda_i^3}\left\{2E_{\rm ISCO} - 2E(\lambda_i) + \lambda_i\left[\left(\frac{dE}{d\lambda}\right)_{\rm ISCO} + \left(\frac{dE}{d\lambda}\right)_{\lambda_i}\right]\right\}\;.
\label{eq:E3def}
\end{eqnarray}
\end{widetext}
We similarly define ${\cal L}^{\rm M2}_2$, ${\cal L}_3$, ${\cal Q}^{\rm M2}$, and ${\cal Q}_3$ to insure that $L_z$ and $Q$ smoothly evolve through the transition.

The lower panel of Fig.\ {\ref{fig:energy_models}} shows both $E^{\rm M1}$ and $E^{\rm M2}$, comparing to the OT00 model shown in the upper panel.  The energy and its slope are each continuous at $\lambda_i$, eliminating the problematic discontinuity.  Model 1 (the green curve) is quite flat across the transition, abruptly becoming constant at $M\lambda \simeq 16.8$, when we switch to the plunge model.  Model 2 (the blue curve) exhibits more curvature in the transition, dipping down so that $E^{\rm M2}(0) = E_{\rm ISCO}$.  This dip means that the end of the transition is less abrupt, changing to the constant value characterizing plunge fairly smoothly.

In Sec.\ {\ref{sec:results}}, we examine worldlines computed using both of our refinements to OT00.  The smooth behavior of Model 2 leads us to slightly prefer this choice to Model 1.  Model 2 is also closer to the spirit of the original Ori-Thorne model, in which various inputs to the transition model are developed as an expansion about their values at the ISCO.  We emphasize, though, that both models are essentially {\it ad hoc} modifications which ensure that no non-physical discontinuities contaminate our worldlines.

It is also worth emphasizing that these refinements to the evolution of $E$, $L_z$, and $Q$ are inconsistent with the form that was assumed in deriving Eq.\ (\ref{eq:ottransgen}).  In principle, one could imagine revisiting that derivation and making a more complicated variant of the transition equation, but we would then lose the benefit of having the universal solution $X(L)$ that describes the transition region.  For now, we have chosen to just live with this internal inconsistency.  As discussed below, there are several {\it ad hoc} choices that must be made to construct these worldlines, and the impact of the other choices appears to be larger than any error incurred by the inconsistency of refinement with Eq.\ (\ref{eq:ottransgen}).  The need for these refinements and the inconsistency that they introduce is yet another reason that it would be good to develop a more rigorously justified model for inspiral and plunge for large mass ratio, inclined inspirals.  In any case, our results indicate that Model 1 and Model 2 produce nearly identical inspiral, transition, and plunge worldlines, consistent with the fact that the terms we add are quite small in numerical magnitude.

\subsubsection{Plunge}

At our chosen final time $L_f$, we freeze the constants $E$, $L_z$, and $Q$ to their values corresponding to $\lambda_f$ that we find using Eq.\ (\ref{eq:LdefGOT}).  The radial motion is then governed by solving the second-order radial geodesic equation, Eq.\ (\ref{eq:minoradial}).  For our initial condition, we use Eqs.\ (\ref{eq:XdefGOT}) and (\ref{eq:LdefGOT}) to set $r_{\rm o} = r(\lambda_f)$, and to find the first derivative $dr_{\rm o}/d\lambda$ at $\lambda_f$.  It is then a simple numerical exercise to integrate this differential equation to compute $r_{\rm o}(\lambda)$ along this plunging trajectory.  We end the plunge when the small body has passed inside the event horizon, i.e.\ when $r_{\rm o} < r_{\rm H} = M + \sqrt{M^2 - a^2}$.

\subsection{Worldline in Boyer-Lindquist time}
\label{sec:WLstep2}

The outcome of the procedure described above is a set of functions $[r_{\rm o}(\lambda), E(\lambda), L_z(\lambda), Q(\lambda)]$ defined on a domain $\lambda_{\rm start} \le \lambda \le \lambda_{\rm end}$.  This set describes the trajectory of the infalling body from large radius through the black hole's event horizon, parameterized in Mino time.  We next convert the parameterization from Mino time to Boyer-Lindquist time, computing the small body's motion in the angles $\theta$ and $\phi$ as we do so.  Boyer-Lindquist coordinate time corresponds to the time that is used by distant observers of the system, so this puts the worldline into a form appropriate for computing measurable quantities.  It also puts the worldline into the form suited for constructing the source function of time-domain BHPT.

The set of equations that we integrate is based on Eqs.\ (\ref{eq:phidot}), (\ref{eq:tdot}), (\ref{eq:geodesiceqnsMino}), and (\ref{eq:dchidlambda}):
\begin{eqnarray}
\frac{d\lambda}{dt} &=& \frac{1}{T(r_{\rm o},\theta)}\;,\quad
\frac{d\chi}{dt} \equiv \frac{d\chi}{d\lambda}\frac{d\lambda}{dt}\;,
\label{eq:dlambdaanddchidt}\\
\frac{d\phi}{dt} &\equiv& \frac{d\phi}{d\tau}\left(\frac{dt}{d\tau}\right)^{-1}
\nonumber\\
&=& \frac{2aMEr_{\rm o} - a^2L_z + \Delta(r_{\rm o})L_z\csc^2\theta}{E(r_{\rm o}^2 + a^2)^2 - 2 a M L_z r_{\rm o} - \Delta(r_{\rm o}) a^2 E \sin^2\theta}\;.
\nonumber\\
\label{eq:dphidt}
\end{eqnarray}
The function $T(r,\theta)$ is defined in Eq.\ (\ref{eq:tdot}), $d\chi/d\lambda$ is defined in Eq.\ (\ref{eq:dchidlambda}), and $\Delta(r) = r^2 - 2Mr + a^2$.

We integrate these equations by stepping evenly in $t$, building up $\lambda(t)$ using Eq.\ (\ref{eq:dlambdaanddchidt}).  We choose initial conditions $t = 0$ at $\lambda = \lambda_{\rm start}$, $\phi = 0$ at $t = 0$, and $\chi = 0$ at $t = 0$.  At each step, we update $r_{\rm o}$, $E$, $L_z$, and $Q$ using the Mino-time worldline we constructed in the previous step.  The result is a set of functions $\lambda(t)$, $\chi(t)$, and $\phi(t)$; we convert $\chi(t)$ into $\theta(t)$ using Eq.\ (\ref{eq:chidef}).  (The parameter $\chi_0$ in this equation allows us to choose different starting values for $\theta$ given a particular choice of $I$.)

Recall that the event horizon is defined by the condition $\Delta(r_{\rm H}) = 0$.  This means that $1/T(r_{\rm o},\theta) \to 0$ as the horizon is approached, and that
\begin{eqnarray}
\frac{d\lambda}{dt} &\to& 0\;,\quad
\frac{d\chi}{dt} \to 0\;,
\label{eq:radialandpolarfreeze}\\
\frac{d\phi}{dt} &\to& \frac{a}{2Mr_{\rm H}} \equiv \Omega_{\rm H}\;
\label{eq:axialfreeze}
\end{eqnarray}
as $r_{\rm o} \to r_{\rm H}$.  As the infalling body approaches the horizon, $\lambda$ and $\chi$ stop evolving as seen by distant observers.  The body freezes at the radial and polar coordinates at which it approaches the horizon, and whirls in axial angle at a constant angular frequency $\Omega_{\rm H}$.

\section{Results}
\label{sec:results}

\subsection{Behavior of our inspiral, transition, and plunge worldlines}

\begin{figure}[h!tbp]
\includegraphics[width=0.47\textwidth]{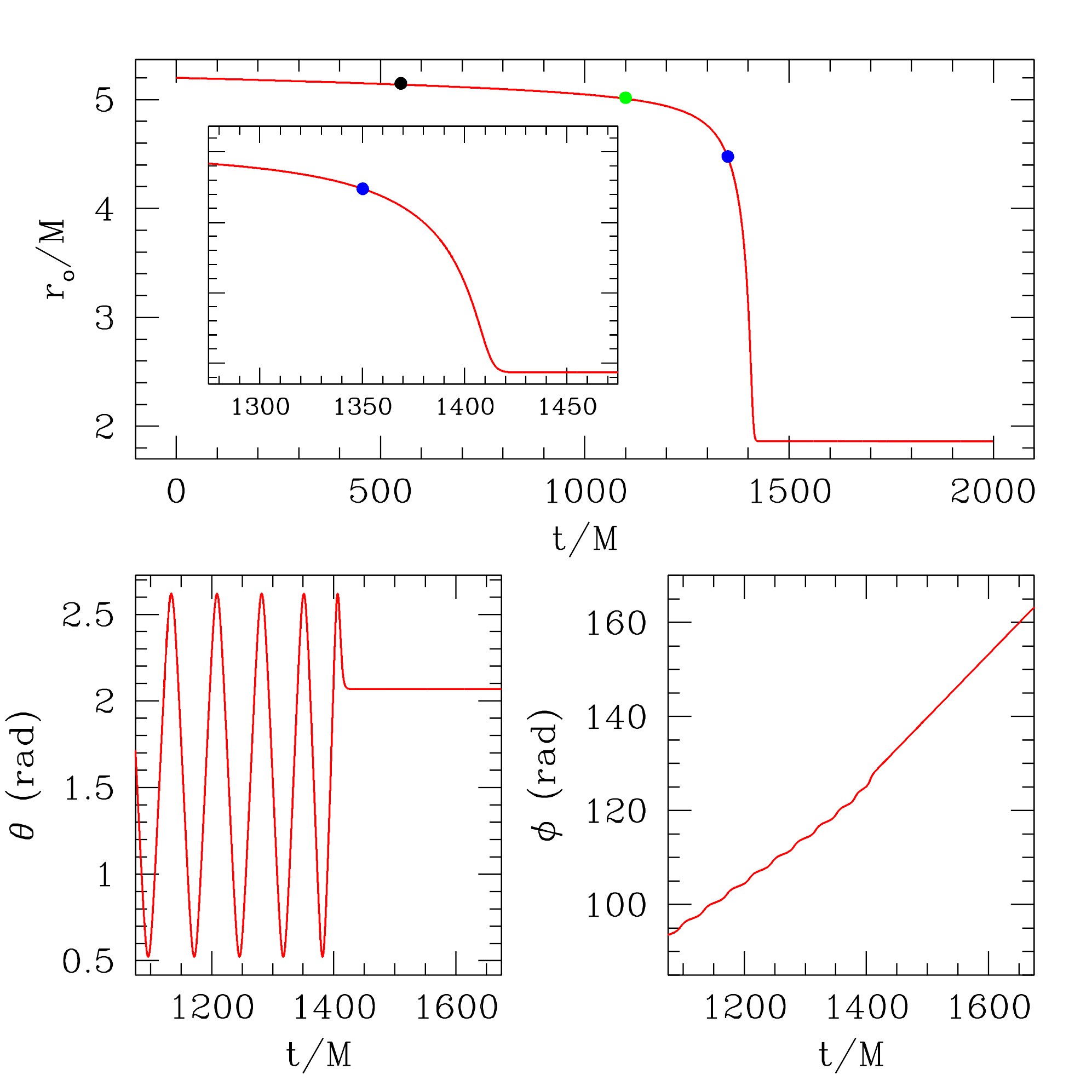}
\caption{A typical example worldline constructed using our generalized Ori-Thorne prescription.  These data describe a system with a mass ratio $\eta = 10^{-4}$ inspiraling with inclination angle $I = 60^\circ$ into a black hole with spin $a = 0.5M$.  Top panel shows the orbital radius as a function of time.  The moment when we switch from adiabatic inspiral to the transition regime motion is marked by the black dot at $t \simeq 550M$; the moment when we switch from the transition to the plunge is marked by the blue dot at $t \simeq 1350M$.  After passing $r_{\rm ISCO}$ (marked by the large green dot), the orbit's infall rapidly accelerates until at $t \simeq 1420M$ its radial motion in these coordinates freezes at the horizon ($r_{\rm H} = 1.866M$); the inset panel zooms onto the motion in the moments as approaches and the freezes to the horizon.  The polar coordinate $\theta$ (lower left panel) likewise oscillates between $30^\circ$ and $120^\circ$ until it also freezes at the horizon at $\theta = \theta_f = 115^\circ$.  The value of $\theta_f$ depends upon the orbit's phase as it enters the plunge, and can be varied using the parameter $\chi_0$ in Eq.\ (\ref{eq:chidef}); see Fig.\ {\ref{fig:chidep}}.  The axial coordinate $\phi$ (lower right panel) steadily accumulates with small oscillations in the rate of accumulation.  The oscillations end when the small body freezes onto the horizon, when the growth of $\phi$ becomes linear, with slope $\Omega_{\rm H} = a/2Mr_{\rm H} = 0.134/M$.}
\label{fig:wlexample}
\end{figure}

Figure {\ref{fig:wlexample}} shows a typical example of the behavior that we see in the worldlines that we construct.  This figure shows the worldline for inspiral and plunge into a black hole with spin $a = 0.5M$ with inclination angle $I = 60^\circ$ at mass ratio $\eta = 10^{-4}$.  The parameters $L_i$ and $L_f$, which as discussed in Sec.\ {\ref{sec:GOT}} mark the beginning and end of the transition epoch, are chosen to have the values $L_i = -3$ and $L_f = 2.5$.  How the worldlines behave when these choices are modified will be discussed momentarily.

The top panel of Fig.\ {\ref{fig:wlexample}} shows how the orbital radius evolves as a function of Boyer-Lindquist time.  For much of the time, $r$ changes very slowly.  The infall becomes significantly faster as the small body approaches $r_{\rm ISCO}$ (marked by the green dot), falling from $r \simeq 4.8M$ to the horizon $r_{\rm H} = 1.866M$ in a time interval of about $100M$.  By $t \simeq 1420M$, the infalling body motion has frozen to the horizon in these coordinates.  The polar angle $\theta$ (lower left panel of Fig.\ {\ref{fig:wlexample}}) oscillates between $30^\circ$ and $120^\circ$ until the small body plunges; it locks to the horizon at $\theta = \theta_f = 115^\circ$.  The final value of $\theta$ depends upon the orbit's phase as it enters the plunge, and can be changed by adjusting the polar phase parameter $\chi_0$ [see Eq.\ (\ref{eq:chidef})].  The axial angle $\phi$ (lower right panel of Fig.\ {\ref{fig:wlexample}}) accumulates over the inspiral and plunge.  The growth of $\phi$ shows oscillations in its rate of accumulation during the inspiral, then shows constant growth at $d\phi/dt = \Omega_{\rm H} = a/2Mr_{\rm H} = 0.134/M$ after the small body has frozen onto the horizon as seen by distant observers.

\begin{figure*}[h!tbp]
\includegraphics[width=0.48\textwidth]{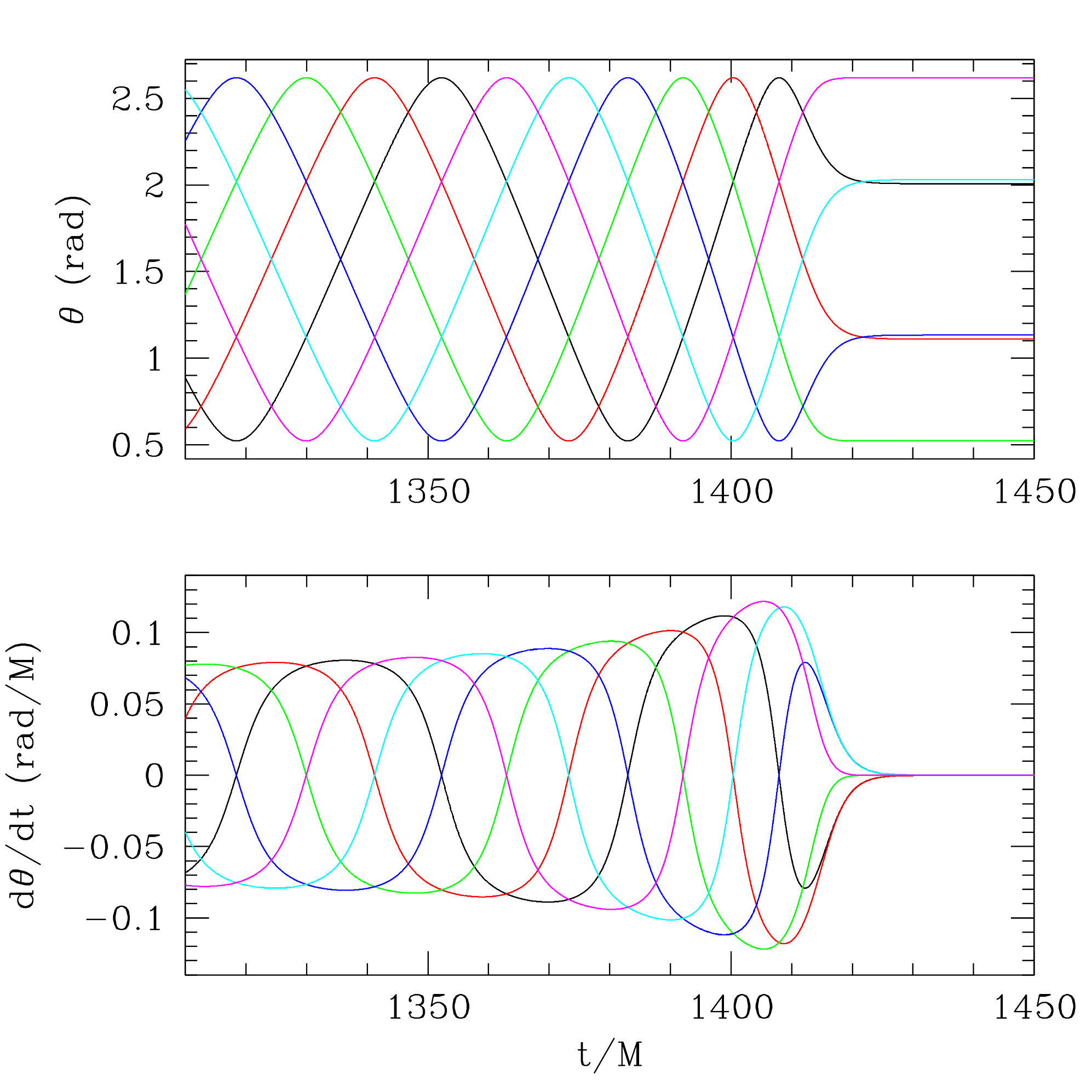}
\includegraphics[width=0.48\textwidth]{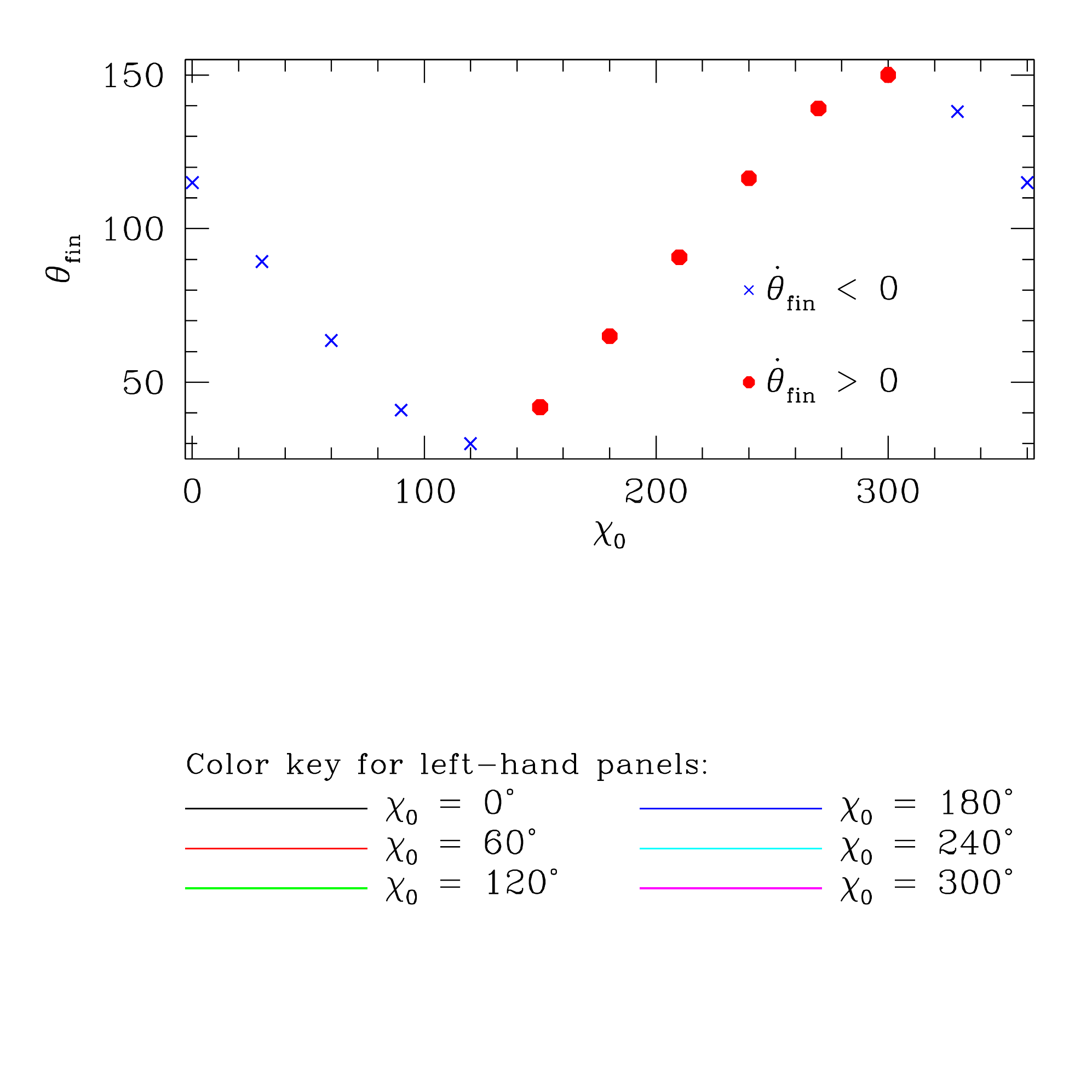}
\caption{Left-hand panels: Behavior of the polar angle versus time for the same system used to make Fig.\ {\ref{fig:wlexample}}, but varying the polar phase $\chi_0$ defined in Eq.\ (\ref{eq:chidef}).  Top left shows the angle $\theta$ versus time; bottom left shows its time derivative.  The top right panel shows how each value of $\chi_0$ maps to a particular value of $\theta_f$ at which the orbit freezes onto the horizon.  Red crosses label orbits for which $d\theta/dt < 0$ as the small body approaches the horizon, blue dots are those for which $d\theta/dt > 0$ at the horizon.  (Bottom right is a color key for the left-hand panels.)  Note that each value of $\theta_f$ corresponds to two values of $\chi_0$, depending upon whether the worldline enters $\theta_f$ with $d\theta/dt > 0$ or with $d\theta/dt < 0$.}
\label{fig:chidep}
\end{figure*}

Figure {\ref{fig:chidep}} explores how varying the polar phase $\chi_0$ changes $\theta_f$, the polar angle at which the worldline freezes onto the horizon.  We show both $\theta$ (top panel) and $d\theta/dt$ (bottom panel) for $\chi_0 \in [0^\circ, 60^\circ, 120^\circ, 180^\circ, 240^\circ, 300^\circ]$.  By varying $\chi_0$, we can select $\theta_f$ to have any value in the range $\theta_{\rm min} \le \theta_f \le \theta_{\rm max}$ [with $\theta_{\rm min/max}$ defined by Eq.\ (\ref{eq:thetaminmaxdefs})].  We also see that each value of $\theta_f$ corresponds to two values of $\chi_0$, depending on whether the last moments of the plunge have $d\theta/dt > 0$ or $d\theta/dt < 0$.  As we will discuss in our companion paper {\cite{lkah}}, we have found these two branches have important implications for the excitation of black hole ringing modes.

\subsection{Robustness of our worldlines versus model parameter choices}

The generalized Ori-Thorne algorithm we have developed requires us to make three {\it ad hoc} choices: which refinement to use to evolve $E$, $L_z$, and $Q$ through the transition, Model 1 [Eq.\ (\ref{eq:energy_model1})] or Model 2 [Eq.\ (\ref{eq:energy_model2})]; the value of $L_i$ at which to change from the adiabatic inspiral to the transition model; and the value of $L_f$ at which to change from the transition model to the plunge.  We have found our generalized algorithm to be essentially completely insensitive to our choice of transition model, at least for $\eta < 10^{-2}$, and to be only slightly sensitive to the choice of $L_i$.  It is, however, sensitive to our choice of $L_f$.  We have found {\cite{lkah}} that this sensitivity does not have any impact on our ability to study black hole excitation, provided we remain within the domain $2.2 \le L_f \le 2.5$, and provided we parameterize our worldlines using the angle $\theta_f$.  Nonetheless, it would be very desirable to develop a method for producing highly inclined inspiral and plunge worldlines that were not sensitive to such an {\it ad hoc} parameterization.

\begin{figure}[h!tpb]
\includegraphics[width=0.46\textwidth]{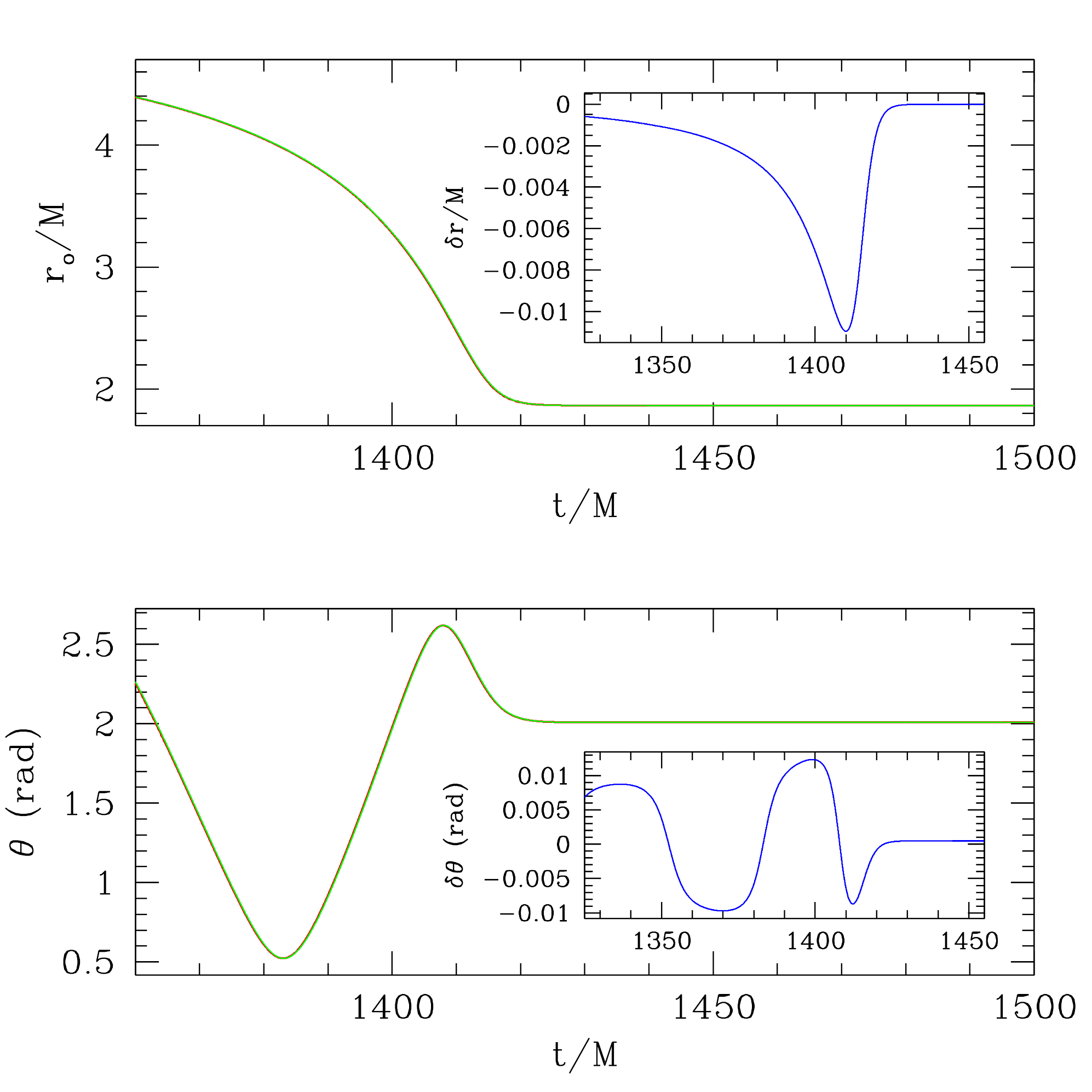}
\caption{Comparing the effect of different values for the transition start parameter $L_i$ (discussed in Sec.\ {\ref{sec:GOT}}).  We examine a set of worldlines constructed by selecting $-5 \le L_i \le -1$.  All choices in this range are consistent with the heuristic guidance laid out in Appendix {\ref{app:heuristic}}.  The top panel shows the resulting radial motion, and the bottom the resulting polar motion.  In both cases, we show worldlines for the choices $L_i = -5$ (red curve) and $L_i = -1$ (green curve).  The difference between the two choices can barely be discerned on this plot; indeed, the green curve (which was plotted second) almost entirely covers the red curves.  In the insets, we show the differences $\delta r \equiv r_{L_i = -5} - r_{L_i = -1}$ and $\delta\theta \equiv \theta_{L_i = -5} - \theta_{L_i = -1}$.  Over this range, the radial trajectories never differ by more than $0.01M$, and the angular motion never differs by more than about $0.012$ radian.  As long as $L_i$ is selected from this range, the worldlines we compute barely differ from one another.}
\label{fig:transstartcomp}
\end{figure}

Figure {\ref{fig:transstartcomp}} shows the radial (top panel) and polar (bottom) motion for two systems that are nearly identical to the one shown in Fig.\ {\ref{fig:wlexample}}, except that we use $L_i = -5$ (red curve) and $L_i = -1$ (green curve).  For both the radial and polar motion, the two models barely differ: the green curves (which were plotted second) almost completely cover the red curves.  The inset in the two panels show the difference in the two models: the radial trajectories never differ by more than $0.01M$, and the angular trajectories never differ by more than about $0.012$ radian.  The angles $\theta_f$ at which the two trajectories freeze to the horizon differ by $2 \times 10^{-4}$ radian.

We have found essentially identical behavior when we examine the impact of $L_i$ on a wide variety of different inspiral and plunge systems (indeed, the differences found for the case shown in Fig.\ \ref{fig:transstartcomp} are somewhat larger than is typical when one considers an ensemble of worldlines).  For all further analysis, we chose $L_i = -3$, in the middle of the range we consider.

Similar results are seen when we compare our two prescriptions for evolution of the integrals of motion through the transition: we see negligible differences between worldlines constructed using Eq.\ (\ref{eq:energy_model1}) and Eq.\  (\ref{eq:energy_model2}).  To see any significant difference between the two prescriptions, we must use $\eta > \mbox{several}\times10^{-2}$, a mass ratio at which first-order BHPT theory is surely not valid.  In all the results we present here and in our companion analysis, we use Model 2 [Eq.\ (\ref{eq:energy_model1})] to evolve $E$, $L_z$, and $Q$ through the transition.

\begin{figure}
\includegraphics[width=0.48\textwidth]{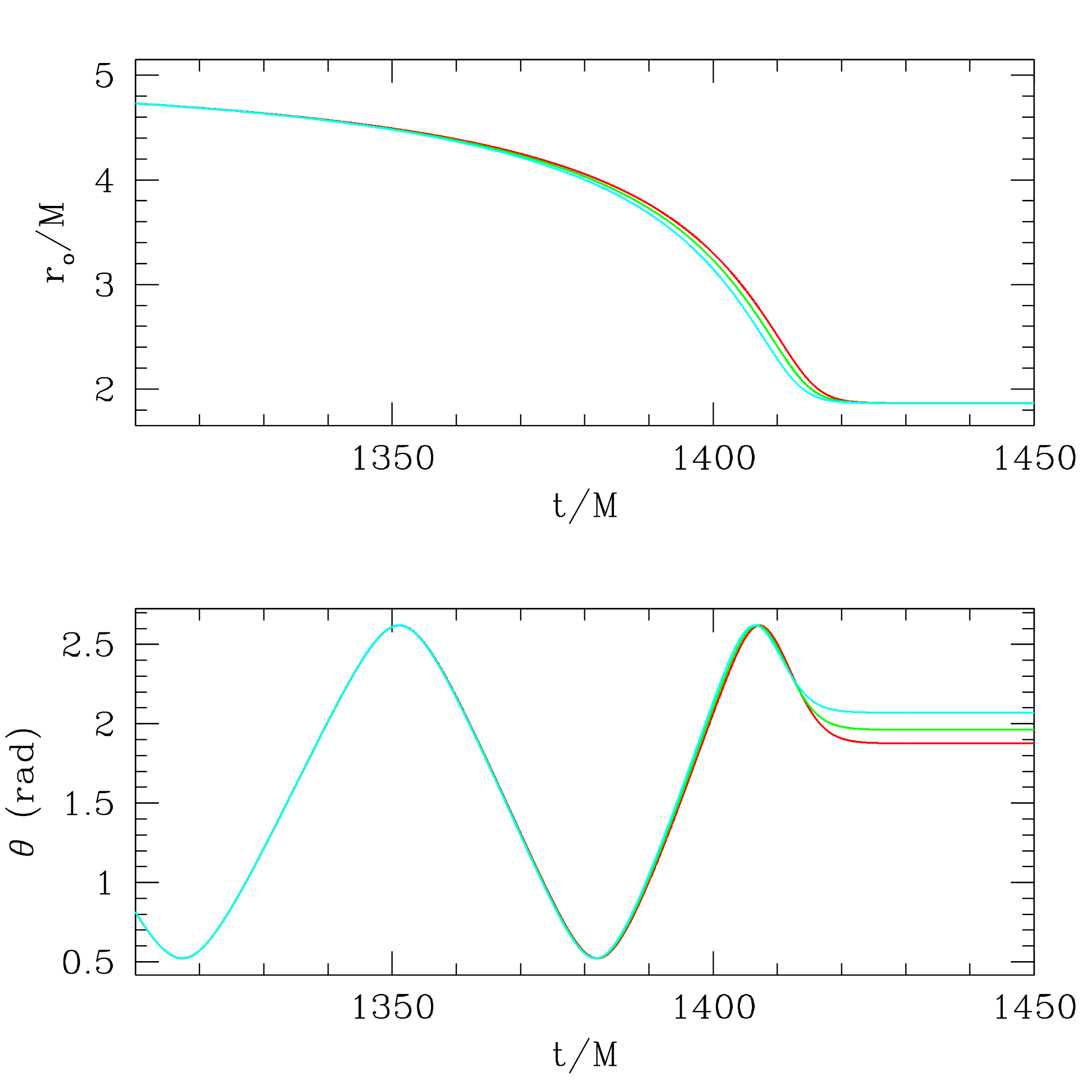}
\caption{Comparing the effect of different values for the transition end parameter $L_f$ (discussed in Sec.\ {\ref{sec:GOT}}).  We show worldlines for three different choices, $L_f = 2.2$ (red curves), $L_f = 2.35$ (green curves), and $L_f = 2.5$ (cyan curves).  In contrast to the situation for $L_i$, the worldlines vary significantly depending on $L_f$: the time at which the infalling body freezes to the horizon varies as $L_f$ is varied, as does the value of $\theta$ at which horizon freezing occurs.  As we discuss in our conclusions, this variation does not affect our ability to study how black hole mode excitation depends on orbit geometry.}
\label{fig:transendcomp}
\end{figure}

Figure {\ref{fig:transendcomp}} again shows the radial and polar motion for a set of systems that are nearly identical to the one shown in Fig.\ {\ref{fig:wlexample}}, but now we use $L_f = 2.2$ (red curves), $L_f = 2.35$ (green curves), and $L_f = 2.5$ (cyan curves).  In this case, we see significant differences in the plunge dynamics for both the radial and polar trajectories, with the different trajectories each ending at different values $\theta_f$.  In the context of how the generalized Ori-Thorne algorithm works, is perhaps not terribly surprising that this difference appears: The curve of $X$ versus $L$ (Fig.\ {\ref{fig:XvsT}}) is quite steep in the range $2.2 \le L \le 2.5$.  Small changes to $L_f$ thus translate to relatively large changes to $X$ and thus to large changes in the values of orbital radius and in the quantities $E$, $L_z$, and $Q$.  (Note that the ch anges we show in Fig. 7 are typical, indeed somewhat on the large side, of those we find when we ex amine an ensemble of worldlines with a wide range of parameters.)

Disturbingly, this implies that the {\it ad hoc} choice of when to end the transition and begin the plunge has a noticeable impact upon the trajectories produced by the generalized Ori-Thorne algorithm.  A set of trajectories with the same initial conditions will fall into the horizon at different values of $\theta_f$, depending on the choice of $L_f$.  However, as we discuss in the Conclusion, work that we present in our companion analysis {\cite{lkah}} shows that this effect has no important impact upon our goal of modeling the spectrum of ringdown GWs produced by misaligned black hole coalescences: we find that two trajectories with the same value of $\theta_f$ produce identical ringdown modes.  The dependence of the inspiral and plunge worldline on how the transition ends is thus not important with respect to our larger goal of characterizing gravitational waves from the late-time black hole coalescence waveform, provided we parameterize these waves using the final polar angle $\theta_f$.

\newpage
\section{Conclusion}
\label{sec:conclude}

We have developed a generalization of Ori and Thorne's algorithm for computing the transition between inspiral and plunge which is applicable to inclined orbits, and have used it to explore the worldlines describing a small body which spirals into a Kerr black hole and then plunges into its horizon.  This prescription uses input from frequency-domain BHPT to accurately describe how an orbit's integrals of motion evolve during its inspiral, up to the vicinity of the ISCO.  Beyond that point, the generalized Ori-Thorne algorithm uses a model that requires three {\it ad hoc} parameter choices to describe how the small body's motion makes the transition from slow inspiral to final plunge: a parameter $L_i$ describing when inspiral ends and transition begins; a parameter $L_f$ describing when transition ends and the plunge begins; and a choice of model for how to evolve the integrals of motion beyond the ISCO.

We have found that the worldlines we produce are robust against two of these three {\it ad hoc} choices.  As long as the mass ratio $\eta < 0.01$, we have found that it makes no difference which of the two models we use to evolve the integrals of motion.  Our use of BHPT requires that $\eta$ be small, so either model should be satisfactory for our purposes.  We slightly prefer one model as being closer to the spirit of the OT00; our refinement adds estimators for higher order corrections describing the system's parameters near the ISCO.  We also find that the choice of when to end the inspiral and begin the transition makes very little difference to our analysis.  We examined a range of physically acceptable values of $L_i$, and found that the worldlines produced by any value in that range differ very little from one another.

The worldlines that we produce are {\it not} robust against the third parameter choice, when to end the transition and begin the plunge.  Following OT00, we take the plunge to be an epoch of the system's evolution in which the orbital integrals stop evolving and the small body follows a free-fall geodesic into the larger black hole.  We examined a range of allowable moments $L_f$ to end the transition and begin the plunge.  We find that the detailed behavior of the final plunge depends noticeably upon this choice.  This means that the moment at which the small body freezes onto the black hole's horizon and the polar angle at which this freezing occurs depends sensitively on an unphysical {\it ad hoc} model parameter, a rather unsatisfactory situation.  For equatorial orbits, the effective one-body framework has been used very successfully to build analytic models for the inspiral and plunge of binary systems, including large mass-ratio systems such as we study here.  This framework does not require the {\it ad hoc} choices that we have had to introduce, further motivating the development of a precessing EOB model that matches onto strong-field Kerr orbits.  It may also be possible to construct the inspiral and plunge worldline self consistently; self force codes have so far demonstrated an ability to compute a self consistent evolution of the small body's orbit and its associated radiation at least for scalar self forces and short evolution times {\cite{diener11}}.  At a minimum, applying such an analysis near the end of inspiral may allow us to calibrate how to make these {\it ad hoc} choices in a manner that yields motion similar to a self-consistent analysis.

For our present purposes, we have fortunately found {\cite{lkah}} that the physics we wish to study with these worldlines is not sensitive to our choice of $L_f$.  Our analysis shows that, given black hole spin $a$, mass ratio $\eta$, and orbit inclination $I$, the amount by which different modes are excited by the final plunge depends on the value of the polar angle value $\theta_f$ at which the small body freezes to the horizon (as seen by distant observers).  We have found that different choices of the parameter $L_f$ produce the same {\it family} of mode excitations.  Two worldlines that share $a$, $\eta$, $I$, and $\theta_f$ excite the same set of black hole modes, even if they use different values of $L_f$.  These worldlines will differ in their {\it initial} phases (they will have different values of $\chi_0$), but as long as $\theta_f$ is the same, they will excite the same ringdown modes.

We are thus confident that we have developed useful tools for exploring how black hole mode excitation depends on the geometry of a binary's inspiral and plunge.  Results describing our exploration of mode excitation are presented in a companion paper {\cite{lkah}}.

\section*{Acknowledgments}

We thank Gaurav Khanna and Halston Lim for many discussions in the course of completing this analysis.  Our work on this problem was supported by NSF Grants PHY-1403261 and PHY-1707549.

\appendix

\section{A heuristic overview of the transition from inspiral to plunge}
\label{app:heuristic}

In this appendix, we develop a heuristic sketch of the transition between inspiral and plunge which makes it possible to estimate the radius at which inspiral ends and the transition begins.

As described in Sec.\ {\ref{sec:kerrcirc}}, circular orbits for a large mass-ratio binary are defined by the conditions $R = 0$, $R' = 0$, where $R$ is the potential-like function defined in Eq.\ (\ref{eq:rdot}), and prime denotes $\partial/\partial r$.  The adiabatic inspiral denotes the period of the binary's evolution in which backreaction is ``slow enough" that the orbit follows that extremum as $R$ changes due to GW emission.

What, precisely, does ``slow enough'' mean?  To understand this better, let us examine qualitatively how the orbit  follows the extremum of $R$.  As the binary emits GWs, the quantities ($E, L_z, Q$) change due to the waves' backreaction.  This in turn changes the function $R$, in particular moving its extremum to smaller radius.  As $R$ evolves, the orbit will become momentarily displaced from its extremum.  In this displaced position, the curvature of $R$ will push the orbit toward the new extremum.  In essence, $R$ provides a restoring force which pushes the orbit toward the slowly evolving location of its extremum.

As long as the backreaction of GWs is not too strong, this restoring force will be able to ``keep up'' with the inward drift of the extremum.  In this case, the system can be modeled as adiabatically moving through a sequence of geodesic orbits.   As the backreaction of GWs becomes stronger, $R$ will at some point begin to change more quickly than the restoring force can respond.  The orbit can then no longer ``keep up'' with the evolving extremum, and so the system can no longer adiabatically track a geodesic sequence.  When this happens, the system has begun its transition to the plunge.

Let us now make this qualitative picture quantitative.  To do this, we must compare two timescales: the time $T_R$ associated with the restoring force from $R$, and the time $T_{\rm GW}$ associated with how quickly the small body's inspiral accelerates as it moves into the black hole's strong field.  We define the system to be in the adiabatic inspiral when $T_{\rm GW} > T_R$ (so that the backreaction of GWs is slower than the restoring action of $R$'s curvature), but say that it has begun the transition to plunge when $T_{\rm GW} < T_R$ (when GWs act more quickly than $R$).

Begin with $T_R$.  For simplicity, we focus on equatorial orbits in this heuristic discussion (although the rough calculation that we sketch can be generalized to non-equatorial orbits), and we use the potential $V_r$ introduced by Ori and Thorne (cf.\ Sec.\ {\ref{sec:OT}}).  Begin with the equation for the radial acceleration:
\begin{equation}
\frac{d^2r}{d\tau^2} = -\frac{1}{2}\frac{\partial V_r}{\partial r}\;.
\label{eq:radialacc_OT}
\end{equation}
Consider an orbit that is slightly perturbed from circular, so that its radius satisfies $r_{\rm o} = r_{\rm circ} + \delta r$.  Using the fact that $\partial V_r/\partial r = 0$ at $r = r_{\rm circ}$, Eq.\ (\ref{eq:radial2}) becomes
\begin{equation}
\frac{d^2(\delta r)}{d\tau^2} + \omega_r^2 \delta r = 0\;,
\label{eq:perturbedcirceq2}
\end{equation}
where
\begin{equation}
\omega_r = \sqrt{\frac{1}{2}\left(\frac{\partial^2V_r}{\partial r^2}\right)}\;.
\end{equation}
This $\omega_r$ is the frequency (conjugate to the orbit's proper time $\tau$) at which the slightly non-circular orbit oscillates about the extremum of the potential $V_r$.  The inverse of this frequency is a timescale characterizing how quickly the restoring force associated with the potential pushes the orbit toward the extremum.  Converting from proper time to Boyer-Lindquist time, we have
\begin{equation}
T_R = \sqrt{\frac{2}{g_{tt}}\left(\frac{\partial^2V_r}{\partial r^2}\right)^{-1}}\;.
\end{equation}
All quantities inside the root are evaluated at $r = r_{\rm circ}$.

Consider next the rate at which the small body inspirals due to GW emission.  We want to compute the inward drift velocity associated with the GW backreaction, and the rate at which that drift accelerates:
\begin{eqnarray}
\left(\frac{dr}{dt}\right)^{\rm GW} &=& \frac{dE/dr}{(dE/dt)^{\rm GW}}\;,
\nonumber\\
\left(\frac{d^2r}{dt^2}\right)^{\rm GW} &=& \left(\frac{dr}{dt}\right)^{\rm GW}\frac{d}{dr}\left(\frac{dr}{dt}\right)^{\rm GW}\;.
\end{eqnarray}
Here, $dE/dr$ is the radial derivative of the orbital energy, given by Eq.\ (\ref{eq:E_eq}) for equatorial orbits, and $(dE/dt)^{\rm GW}$ is the rate at which the energy evolves due to GW emission.  For circular orbits, this quantity can be written
\begin{equation}
\left(\frac{dE}{dt}\right)^{\rm GW} = -\frac{64}{5}\eta^2 \frac{\dot{\cal E}}{r^5}\;.
\label{eq:dEdt}
\end{equation}
The right-hand side of (\ref{eq:dEdt}) is the leading quadrupole GW emission, modulo a correction factor $\dot{\cal E}$ which is computed numerically and varies very slowly with $r$.  For $a = 0$, $\dot{\cal E} \simeq 1.14$ in the vicinity of the ISCO; Table I of OT00 lists its value for a range of black hole spins.  The GW timescale we seek is the ratio
\begin{equation}
T_{\rm GW} = \frac{(dr/dt)^{\rm GW}}{\left(d^2r/dt^2\right)^{\rm GW}}
= \left[\frac{d}{dr}\left(\frac{dr}{dt}\right)^{\rm GW}\right]^{-1}\;.
\end{equation}

Using Eqs.\ (\ref{eq:E_eq}) and (\ref{eq:dEdt}), we compute $T_{\rm GW}$ and $T_R$ and compare.  The result is quite simple for $a = 0$:
\begin{equation}
\frac{T_{\rm GW}}{T_R} = \frac{5}{192\dot{\cal E}\eta}\frac{r_{\rm o}^3(r_{\rm o} - 6M)^{5/2}}{(r_{\rm o}-3M)(2r_{\rm o}^2 - 17M r_{\rm o} + 42M^2)}\;.
\label{eq:timescaleratio}
\end{equation}
Similar results can be found for general black hole spin and for non-equatorial orbits, though the formulas are much more complicated.

As discussed above, the system evolves adiabatically when $T_{\rm GW} > T_R$, and is no longer adiabatic when $T_{\rm GW} < T_R$.  We wish to estimate the orbital radius at which adiabaticity is beginning to end, so let us solve for the value of $r_{\rm o}$ at which $T_{\rm GW}/T_R = A$, where $A \sim 1 - 10$.  Putting $r_{\rm o} = 6M + x$, expanding in $x$, and solving for $T_{\rm GW}/T_R = A$, we find
\begin{equation}
x = \frac{4M\dot{\cal E}^{2/5}}{5^{2/5}}A^{2/5}\eta^{2/5}
\simeq 0.14M\times \left(\frac{A}{10}\right)^{2/5}\left(\frac{\eta}{10^{-4}}\right)^{2/5}\;.
\label{eq:transition_heuristic}
\end{equation}
The final numbers we present use the value $\dot{\cal E} \simeq 1.14$ appropriate for $a = 0$.  For $a = 0$ and $\eta = 10^{-4}$, $x = 0.14M$ corresponds to ending the inspiral and beginning the transition at $L_i = -3.6$, right in the domain $-4 \le L_i \le -1.4$ that we argued was appropriate in Sec.\ {\ref{sec:worldline}}.

Although we have presented numbers only for the case of equatorial Schwarzschild inspiral, it is straightforward to generalize to more generic situations.  The key lesson of this analysis is that the inspiral ends at a radius that scales with the system's mass ratio to the $2/5$ power, and is quite close to the ISCO for geodesic orbits.  Both this scaling and the value we find comports with the choices we advocate for switching from inspiral to transition as discussed in Sec.\ {\ref{sec:worldline}}.

\end{document}